\documentclass[]{aa}  

\usepackage{graphicx}
\usepackage{txfonts}
\usepackage{color}
\begin{document}

   \title{Initial Mass Function Variations Cannot Explain the Ionizing Spectrum of Low Metallicity Starbursts}

  % \subtitle{I. Overviewing the $\kappa$-mechanism}
\titlerunning{IMF effects on Ionizing Spectra}

   \author{E. R. Stanway
          \inst{1}
          \and
          J. J. Eldridge\inst{2}%\fnmsep\thanks{Just to show the usage of the elements in the author field}
          }

   \institute{Physics Deparment, University of Warwick, Gibbet Hill Road, Coventry, CV4 7AL, UK\\
              \email{e.r.stanway@warwick.ac.uk}
         \and
             Department of Physics, University of Auckland, Private Bag 92019, Auckland, New Zealand\\
             \email{j.eldridge@auckland.ac.nz}
             %\thanks{The university of heaven temporarily does not accept e-mails}
             }

   \date{Received 2nd Oct 2018; accepted 8th Nov 2018}

% \abstract{}{}{}{}{} 
% 5 {} token are mandatory
 
  \abstract
  % context heading (optional)
  % {} leave it empty if necessary  
   {}
  % aims heading (mandatory)
   {Observations of both galaxies in the distant Universe and local starbursts are showing increasing evidence for very hard ionizing spectra that stellar population synthesis models struggle to reproduce. Here we explore the effects of the assumed stellar initial mass function (IMF) on the ionizing photon output of young populations at wavelengths below key ionization energy thresholds.}
  % methods heading (mandatory)
   {We use a custom set of binary population and spectral synthesis (BPASS) models to explore the effects of IMF assumptions as a function of metallicity, IMF slope, upper mass limit, IMF power law break mass and sampling.}
  % results heading (mandatory)
   {We find that while the flux capable of ionizing hydrogen is only weakly dependent on IMF parameters, the photon flux responsible for the He\,II and O\,VI lines is far more sensitive to assumptions. In our current models this flux arises primarily from helium and Wolf-Rayet stars which have partially or fully lost their hydrogen envelopes. The timescales for formation and evolution of both Wolf Rayet stars and helium dwarfs, and hence inferred population age, are affected by choice of model IMF. Even the most extreme IMFs cannot reproduce the He\,II ionizing flux observed in some high redshift galaxies, suggesting a source other than stellar photospheres.}
  % conclusions heading (optional), leave it empty if necessary 
   {We caution that detailed interpretation of features in an individual galaxy spectrum is inevitably going to be subject to uncertainties in the IMF of its contributing starbursts. We remind the community that the initial mass function is fundamentally a statistical construct, and that stellar population synthesis models are most effective when considering entire galaxy populations rather than individual objects.}


   \maketitle
%
%________________________________________________________________

\section{Introduction}\label{sec:intrp}

Ionizing photons, those emitted at wavelengths shortwards of the hydrogen Lyman Limit at 912\AA, have played a key role in shaping the evolution of the intergalactic medium and the galaxies which form and interact within it. In the local Universe, the emission of such photons originates from a mixture of sources. Amongst these, two are most prominent: sources powered by accretion onto compact objects and supermassive black holes, and the emission from the photospheres of hot, massive stars.
In the distant Universe, massive active galactic nuclei (AGN) are rarer. Their number density drops far more rapidly than that of star forming galaxies and hence star formation dominates the ionizing photon background \citep{2015ApJ...802L..19R}. Lyman continuum (hydrogen-ionizing) photons escaping from star forming galaxies were responsible for the process of ionizing the intergalactic medium at redshifts of $z\gtrsim6$, and also have an important impact on the thermal conditions within, and emission spectra observed from, the galaxies themselves. As a result there has been strong interest in recent years both in the ways in which ionizing photons leak from their origin galaxies, and in how these photons are produced in the first place \citep[e.g.][]{2018MNRAS.479..994R,2016MNRAS.456..485S,2016MNRAS.458L...6W, 2015MNRAS.453..960M,2016MNRAS.459.3614M,2013ApJ...765...47N}.

An important development in this field has been the development of population synthesis models which account for the effects of binary interactions on stellar evolution pathways. When these are combined with stellar atmosphere models describing the spectral energy distribution of individual stars, the resulting spectral synthesis yields predictions that can be compared to the observed spectral properties of stellar populations \citep{2009MNRAS.400.1019E,2012MNRAS.419..479E,2016MNRAS.456..485S}.0
This has led to an increasing recognition of the importance of massive stars in young stellar populations for reionization, twinned with an appreciation of the uncertainties in their evolution. In particular, both rotation \citep{2015ApJ...800...97T} and stellar multiplicity \citep{2012Sci...337..444S,2017PASA...34...58E} have a strong effect on massive star evolution, and the impact of stellar binaries or multiples may be still more important at low metallicities \citep{2018arXiv180802116M}. Binary star evolution can both boost the ionizing photon flux from a stellar population \citep{2016MNRAS.456..485S} and extend its lifetime \citep{2014MNRAS.444.3466S,2018MNRAS.477..904X}. Crucially this latter property may allow for increased ionizing photon escape fractions, as supernovae from the most massive stars blow channels through the circumstellar and circumgalactic medium, allowing ionizing photon leakage \citep{2018MNRAS.479..994R}.

Improved spectroscopy has made analysis of the ionizing spectra of distant galaxies possible in recent years, primarily through study of those ionizing photons reprocessed by nebular gas in the interstellar environment. With the detection of multiple, high signal-to-noise absorption and emission features in galaxies at $z\sim2-4$ \citep[e.g.][]{2018ApJ...860...75D,2016ApJ...826..159S,2017A&A...608A...4M}, and even higher redshifts in rare cases \citep[e.g.][]{2017MNRAS.464..469S}, it is now possible to construct photoionization models which yield insight into the physical conditions in these star-forming systems. While there is always some degeneracy between assumed gas conditions and the spectrum of the ionizing potential, a number of distant systems show evidence for the presence of a very hard ionizing radiation field. This is observed both in (rest-frame) optical line ratios such as [O\,III\,$\lambda$5007\AA]/[O\,II\,$\lambda$3727\AA] or [O\,III\,$\lambda$5007\AA]/H\,$\beta$\,$\lambda$4681\AA,  and directly in the strength of nebular emission lines including C\,III]\,$\lambda$\,1909\AA\ (which requires a 48\,eV ionization potential) and He\,II\,$\lambda$\,1640\AA\ (at 54.4\,eV) in the ultraviolet \citep{2018ApJ...860...75D,2015ApJ...814L...6R,2018ApJ...862L..10E,2010ApJ...719.1168E,2018ApJ...859..164B}. Interestingly, the class of low redshift extreme emission line galaxies and Lyman-continuum leakers identified as analogues to galaxies in the distant Universe also show hard ionizing spectra \cite[e.g.][]{2018MNRAS.478.4851I}.

Given the association of these hard spectra with intensely star forming galaxies and the absence of clear indications for AGN, a natural question to ask is whether the photospheric emission from stars can provide sufficient ionizing photons, with appropriate energies. The early indications have indicated mixed answers to this question. While stellar populations including interacting binaries have performed well in reproducing the mean properties of galaxies in stacks of high redshift spectra \citep{2016ApJ...826..159S}, and certainly produce a harder radiation field than a single star population \citep{2012MNRAS.419..479E,2017PASA...34...58E,2018MNRAS.477..904X} and sustain this for significantly longer timescales \citep{2014MNRAS.444.3466S,2016MNRAS.456..485S}, reproducing the highest ionization potential lines remains a challenge. In particular, He\,II emission (in either the ultraviolet $\lambda$1640\AA\ or optical $\lambda$4686\AA\ lines) appears to be consistently underestimated given default stellar population synthesis model assumptions \citep[e.g.][]{2018ApJ...862L..10E, 2018ApJ...859..164B}. 

Evolutionary population and spectral synthesis - the complex process of creating a model spectrum for a stellar system of known age, mass and metallicity based on a combination of individual stellar models - involves a number of approximations and assumptions. At its simplest, it requires a grid of stellar evolution models, a grid of stellar atmosphere models to predict observable properties of a star at given temperature and surface gravity, and a function which provides the weighting with which to combine models of different initial mass. More complex cases may include evolutionary tracks which include the effects of stellar rotation, Roche-lobe overflow, common envelope and other binary interactions, and expanded grids of atmosphere models which account for surface composition and stellar winds. For the short-lived massive stars that dominate the ionizing spectrum, the binary fraction is close to unity and the effects of binary interactions can be dramatic \citep[see e.g.][]{2004NewAR..48..861D,2012Sci...337..444S,2017PASA...34....1D,2017PASA...34...58E}. Including these pathways, given an assumed set of mass-dependent binary parameter (period, mass ratio) distributions, is thus essential to accurately predict the ionizing spectrum of high redshift galaxies.

However a more fundamental assumption than the binary parameter assumptions is the underlying initial mass function (IMF). This describes the probability distribution of stellar masses (in the binary case, of primary star masses) at zero age in a starburst, before stellar evolution has any significant effect. For many years, the assumption of a `universal' mass function has been used to simplify modelling of galaxies. This is usually described as a power law function in the form $dN/dM \propto M^{\alpha}$ where N is the number of stars in a population at a given mass M and the function extends between 1 and 100\,M$_\odot$. The most commonly used prescription, based on that of \citet{1955ApJ...121..161S}, uses $\alpha=-2.35$. However, it has become increasingly clear in recent years that this oversimplification is untenable. As the extensive reviews in \citet{2018arXiv180709949H} and \citet{2018arXiv180610605K} discuss in detail, not only is it necessary to break the power law to avoid overpredicting sub-Solar mass stars, but the power law slope at high masses is also highly uncertain. There is evidence for local variations in the IMF and potentially also some metallicity dependence. Studies of the massive ($M>8$\,M$_\odot$) and very massive ($M>100$\,M$_\odot$) star population in the 30\,Doradus region have suggested that the initial mass function in the $M>30\,$M$_\odot$ regime may be as shallow as $\alpha=-1.9^{+0.37}_{-0.26}$ \citep{2018arXiv180703821S,2018Sci...359...69S}. Indirect measurements of the initial mass function in the extragalactic regime also suggest a shallow IMF may be appropriate \citep[e.g.][]{2008MNRAS.391..363W}, and similar shallow functions have been seen in individual local massive galaxies such as NGC\,1399 \citep{2018MNRAS.479.2443V}, while other studies have suggested steeper, `bottom heavy' initial mass functions in elliptical galaxies \citep[albeit in the low stellar mass regime, ][]{2018MNRAS.477.3954P}. 

Determining the effects of the plausible range of IMF variation based on local observations is crucial to deciding whether an additional source of hard ionizing photons is necessary to explain the observed results from intense starbursts. A shallow IMF slope will lead to a population that is `top-heavy', i.e. has more massive stars per unit total starburst mass than expected from a `Salpeter'-like slope. A similar, although subtly different, uncertainty arises from the upper mass cut-off in a stellar population model. While most IMFs do not consider stars with $M>100$\,M$_\odot$, identification of the very massive star population in the Large Magellanic Cloud \citep{2016MNRAS.458..624C} has led to suggestions that the IMF may extend as far as $300$\,M$_\odot$ in intense starburst regions. Given the dependence of the hard ultraviolet spectrum of a population on its hottest stars, the inclusion or neglect of even a small population of very massive stars can have a dramatic effect on the photons capable of ionizing not only hydrogen but also helium. It is thus both timely and necessary to explore the photon flux generated above both 13.6\,eV and 54.4\,eV by young stellar populations, as a function not only of metallicity but also of the assumed initial mass function in a binary stellar population synthesis model.

Is it also instructive to consider the ionizing spectrum at still higher energies. The far-ultraviolet O\,VI $\lambda$1032,1038\AA\ doublet (with an ionization potential of 138\,eV) is frequently seen in absorption in the halos around low redshift star forming galaxies, and has thus been identified both as sensitive to specific star formation rate in galaxies and as a potential tracer of reservoirs of warm ionized gas in the circumgalactic environment \citep{2011Sci...334..948T}. In this case the primary ionization source is the extragalactic ionizing background, which is largely powered in turn by emission from active galactic nuclei, while the association with star formation is provided by the necessity for large scale starburst-driven outflows to account for the density, velocity profile and composition of the absorbing gas \citep[e.g.][]{2015ApJ...811..132M}. Indeed ionization powered by accretion onto AGN, whether indirectly through the extragalactic background, or directly through short-duty-cycle bursts of activity within the galaxies in question, can likely account for many low redshift O\,VI absorption features \citep[e.g.][]{2008ApJS..177...39T,2018MNRAS.474.4740O,2018MNRAS.477..450N}.

However there are signs that this interpretation may not apply universally. Analysis of O\,VI absorbers in the haloes of L$^\ast$ galaxies suggests that ionization by the extragalactic background is strongly disfavoured, and either additional local sources of photoionization or mechanisms such as collisional or shock excitation are required \citep{2016ApJ...833...54W,2015MNRAS.450.2067T}. Observations at high redshift, where the extragalactic background is softer due to evolution in the AGN luminosity function, again imply additional sources of hard photons or collisional mechanisms \citep[][]{2018MNRAS.474.1688C}, while there are also now a handful of examples of intense starbursts which show the ultraviolet O\,VI lines in {\it emission} \citep[e.g.][]{2016ApJ...828...49H}. 

There is general consensus that photoionization models driven by emission from hot stars are unable to produce sufficient photons at these energies to explain observations. Nonetheless, the hints of hard ionizing spectra seen in distant objects suggest that we should not rule out this possibility without exploring the limits of the high mass stellar population given plausible IMF models. While observations of the O\,VI doublet remain challenging at both high and low redshift, it nonetheless provides a useful metric for the hardness of the ionizing spectrum and so will also be considered here.

In this paper, we use a custom set of binary population and spectral synthesis models to explore the effects of initial mass function assumptions on the hard ionizing spectra produced by stellar population synthesis as a function of metallicity. In section \ref{sec:method} we introduce our methodology. In section \ref{sec:slope} we explore the effects of assumed IMF slope and the upper mass cut-off in the massive star population. In section \ref{sec:doublebreak} we introduce a second break in the IMF to shallow slopes at very high mass. In section \ref{sec:mmax} we attempt to quantify the effects of total starburst mass on ionizing photon flux. Finally in section \ref{sec:disc} we discuss the implications of this study and summarise our conclusions in section \ref{sec:conc}.

%________________________________________________________________

\section{Method}\label{sec:method}

\begin{figure}
    \centering
    \includegraphics[width=0.48\textwidth]{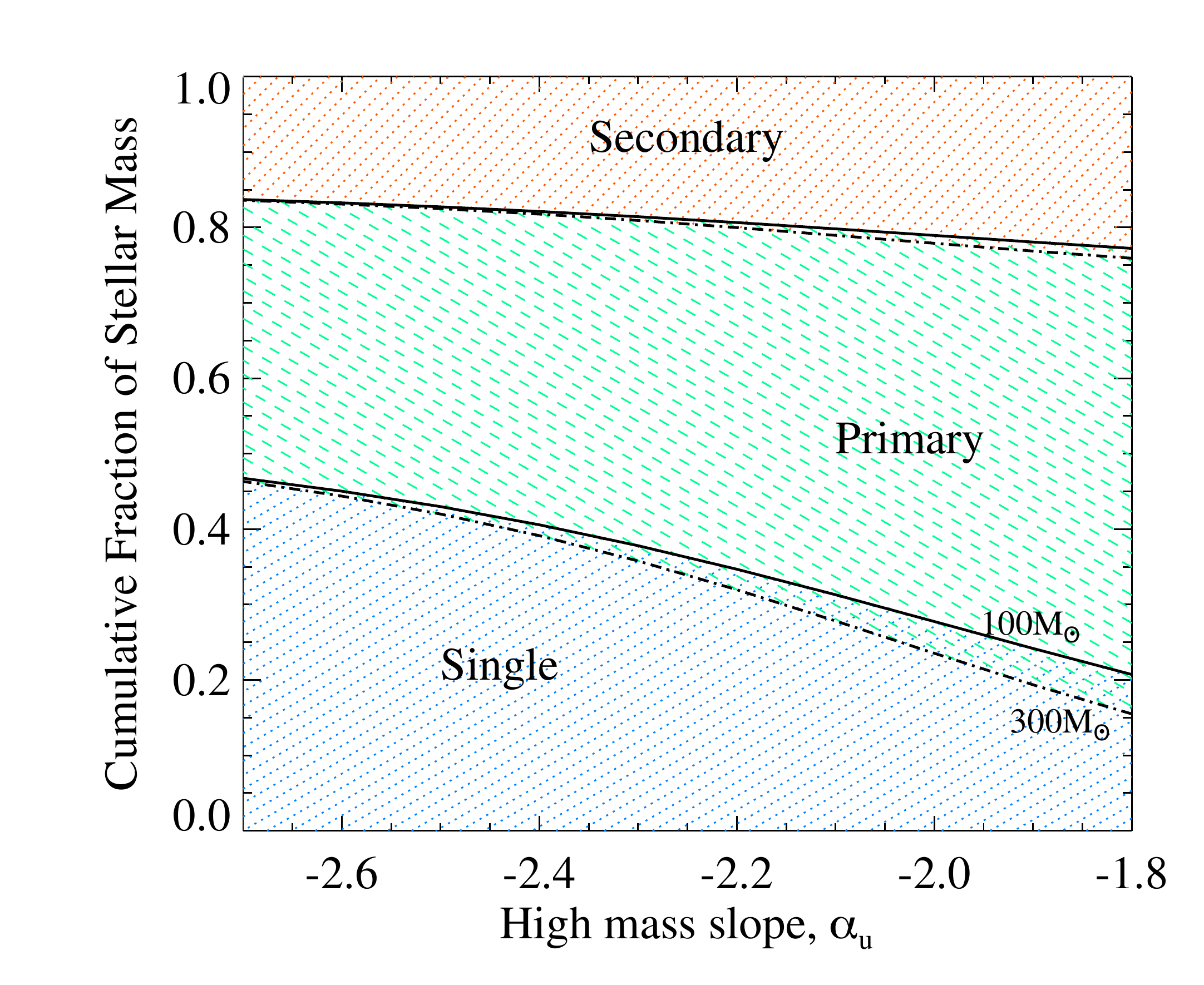}
    \caption{The effect of IMF slope on binary star fraction in a population of total mass $10^6$\,M$_\odot$, given the stellar mass-dependent binary fractions implemented in BPASS v2.2. The lower region between the x-axis and the first pair of lines indicates the fraction of the total stellar mass for a given IMF slope $\alpha_u$ which is assigned to single star models, the central region the mass fraction in primary star models and the upper region the mass fraction assigned to their less massive secondary companions. Regions between each pair of lines (with hatching in two styles) switch between stellar model types depending on the IMF. Models assume a single power law break ($\alpha_m=\alpha_l$), The upper cut-off mass, $M_u$, is shown by line style (100\,$M_\odot$ - solid; 300 - dot-dash). The curves for $M_u=150$ \& 200\,$M_\odot$ lie between these curves.}
    \label{fig:binfrac}
\end{figure}

\subsection{Population Synthesis}\label{sec:popsynth}

The BPASS v2.2 data release \citep{2018MNRAS.479...75S} includes calculation of results for nine initial mass function prescriptions. Of these, one is an unbroken power law and two follow the low mass cut-off prescription of Chabrier (2003) with different upper mass cut-offs. The remainder comprise three pairs of broken power law models with an initial mass function slope of -1.3 between 0.1 and 0.5 M$_\odot$ and -2.0, -2.35 or -2.7 between 0.5\,M$_\odot$ and an upper mass cut-off of either 100\,M$_\odot$ or 300\,M$_\odot$. Calculating such a range of models already represents a substantial investment of resources, but nonetheless this sparse model grid is inadequate to fully explore the effects of initial mass function on the ionizing spectrum.

In order to make such an exploration tractable we modify the BPASS population synthesis to truncate populations at an age of 1 Gyr. This removes the utility of the models as predictors of old stellar populations and astrophysical transient rates, but does substantially speed up the calculation of models with different input initial mass function parameters. We also limit our analysis to a subset of five metallicities, namely $Z=10^{-5}, 10^{-4}, 0.001, 0.010$ and $0.020$, where $Z_\odot=0.020$ is the metal mass fraction of our Sun.  Apart from this, we make no changes to the underlying BPASS v2.2 assumptions (e.g. stellar models, atmosphere models, mass-dependent binary parameters, mass loss rates, mass transfer prescription or synthesis procedure).

For the purposes of this analysis we adopt a twice-broken power law for the initial mass function, of the form:
\begin{equation*}
 \frac{dN(M)}{dM} = A_l\int_{M_0}^{M_l} M^{\alpha_l} dM + A_m\int_{M_l}^{M_m}  M^{\alpha_m} dM + A_u\int_{M_m}^{M_u}  M^{\alpha_u} dM\,,
 \end{equation*}
where $\alpha_l$ is the low mass slope between $M_0$ and lower break mass $M_l$, $\alpha_m$ is the intermediate mass slope valid between $M_l$ and intermediate break mass $M_m$ and $\alpha_u$ is the upper mass slope which is effective between $M_m$ the upper cut-off mass limit $M_u$. The constants $A_l$, $A_m$ and $A_u$ ensure the function is continuous and normalised to yield a total population of $10^6$\,M$_\odot$. In the case where $\alpha_m$=$\alpha_u$ the break mass $M_m$ becomes arbitrary and the form collapses to a simple broken power law.

In table \ref{tab:input_imfs} we summarise both the standard BPASS IMFs and the parameter ranges explored in this work, according to this formulation. In all cases we fix the lower mass limit and power law slope. The stellar binary fraction implemented in BPASS v2.2 is mass-dependent \citep[based on that of][]{2017ApJS..230...15M}. Hence a consequence of varying the IMF is that the  mass fraction of single stars versus binaries (primary and secondary models) is also IMF dependent. In figure \ref{fig:binfrac} we show how the single star fraction varies as a function of both IMF slope and upper mass cut-off. Note that this is independent of metallicity in the current implementation.

\begin{table*}[]
    \centering
    \begin{tabular}{ccccccc}
       $M_0$ / M$_\odot$ & $\alpha_l$ & $M_l$ & $\alpha_m$ & $M_m$ / M$_\odot$ & $\alpha_u$ & $M_u$ / M$_\odot$ \\
     \hline\hline
     0.1 &  -1.30   & 0.5 &  -- & -- &     -2.00 & [100, 300] \\ 
     0.1 &  -1.30   & 0.5 &  -- & -- &    -2.35 &  [100, 300] \\ 
     0.1 &  -1.30   & 0.5 &  -- & -- &    -2.70 &  [100, 300] \\
     0.1 &  -2.35   & 0.5 &  -- & -- &    -2.35 &  100 \\ 
     0.1 & C03 & 1.0 &  -- & -- &    -2.30 &  [100, 300]  \\
      \hline
     0.1 & -1.30    & 0.5 &  -- & -- &   -2.35 & [150, 200, 250] \\
     0.1 & -1.30    & 0.5 &  -- & -- &  [-1.8,-1.9,...-2.6, -2.7] & [100, 150, 200, 300] \\
     0.1 & -1.30    & 0.5 &  -2.35 & [1,5,10,15,20,30,50] &  -2.00 & [100, 150, 200, 300]\,M$_\odot$ \\
     0.1 & -1.30    & 0.5 &  -- & -- &   -2.35 & 300\,M$_\odot$ with prob. cut-off   \\
         & &  & & & &  for $M_\mathrm{tot}$=[1,3,7]$\times$[10$^{2,3,4,5,6,7,8}$]\,M$_\odot$\\
    \end{tabular}
    \caption{Stellar initial mass functions used in BPASS v2.2 (above line) and the extended parameter grid explored in this work. `C03' indicates an exponential cut-off of the form proposed by Chabrier (2003). The total starburst mass dependent probability cut-off in the final model grid is explained in section \ref{sec:mmax}.}
    \label{tab:input_imfs}
\end{table*}

%______
\subsection{Ionizing Spectrum parameterisation}

We define the photon flux generated by each model as:
\begin{equation*}
\hspace{0.5cm}   Q(X) = \int^{\lambda_c}_{1\AA} \frac{\lambda\,L_\lambda}{h\,c} d\lambda\,,
\end{equation*}

where the critical wavelength $\lambda_c = 912$\,\AA\  for Q(H\,I), $227.9$\,\AA\  for Q(He\,II) and $89.79$\,\AA\  for Q(O\,VI), equivalent to the relevant ionization potentials of 13.6, 54.42 and 138.12\,eV respectively. The resultant fluxes are quoted in units of emitted photons per second.

These are calculated at each time step, for each output spectrum generated by our stellar population synthesis models. We  consider photon fluxes rather than directly studying emission or absorption line profiles in the spectra for two reasons: 

Firstly, the flux in spectral features associated with these transitions (e.g. Ly$\alpha$\,$\lambda$\,1216\AA, H$\alpha$\,$\lambda$\,6563\AA, He\,II\,$\lambda$\,1640\AA, He\,II\,$\lambda$\,4686\AA, O\,VI\,$\lambda$\,1035\,AA) is likely to be modified by absorption and re-emission in the nebular gas surrounding young star forming regions. Interpretation of line strengths thus requires an extensive set of assumptions regarding the conditions, covering fraction and composition of the nebular gas screen. 

Secondly, we wish to avoid possible uncertainties introduced into stellar absorption lines by the modelling of stellar atmospheres for rare subgroups within the stellar population. 
%BPASS stellar models predict the luminosity, temperature, surface gravity and surface gravity evolution of the stars in each population modelled.  These must be joined with stellar atmosphere models in order to generate an output spectrum. 
%To do so, for the full range of stellar evolution and binary products identified within BPASS requires the combination of several different libraries of publically-available atmosphere models, initially generated at different spectral resolutions and compositions. At the high temperatures required to generate ionizing photons, stellar atmospheres can be approximated as a black-body emission curve, meaning that the total flux shortwards of a given critical wavelength is largely independent of the atmosphere model selected. However, the emission and absorption features imprinted on the output spectrum will be dependent on the assumptions implicit in the atmosphere models used. In light of the populations studied here we note particularly that 
BPASS implements a custom grid of OB star atmospheres that are required by binary interaction products \citep{2017PASA...34...58E} and also the Potsdam Wolf-Rayet \citep[PoWR,\footnote{ \texttt{www.astro.physik.uni-potsdam.de/}$\sim$\texttt{wrh/PoWR/powrgrid1.php}}][]{2015A&A...577A..13S} atmospheres for stripped helium stars. The PoWR models are extrapolated to lower metallicities than they were created at, which may lead to overprediction in certain metal lines, and modification of the detailed strengths of lines in hydrogen and helium. Due to technical constraints, the current version of BPASS also does not make use of the helium dwarf atmospheres generated by \citet{2017A&A...608A..11G,2018A&A...615A..78G} and these may have a substantial impact on the strength of some emission lines. % which, in our implementation, will likely be dominated by Wolf-Rayet atmospheres. 

As a result, we opt to report directly the ionizing flux generated by a starburst rather than studying the detailed outputs that result from our spectral synthesis.

%__________________________________________________________

\section{Effects of IMF Slope and Upper Mass Limit}\label{sec:slope}

\begin{figure*}
    \centering
    \includegraphics[width=0.45\textwidth]{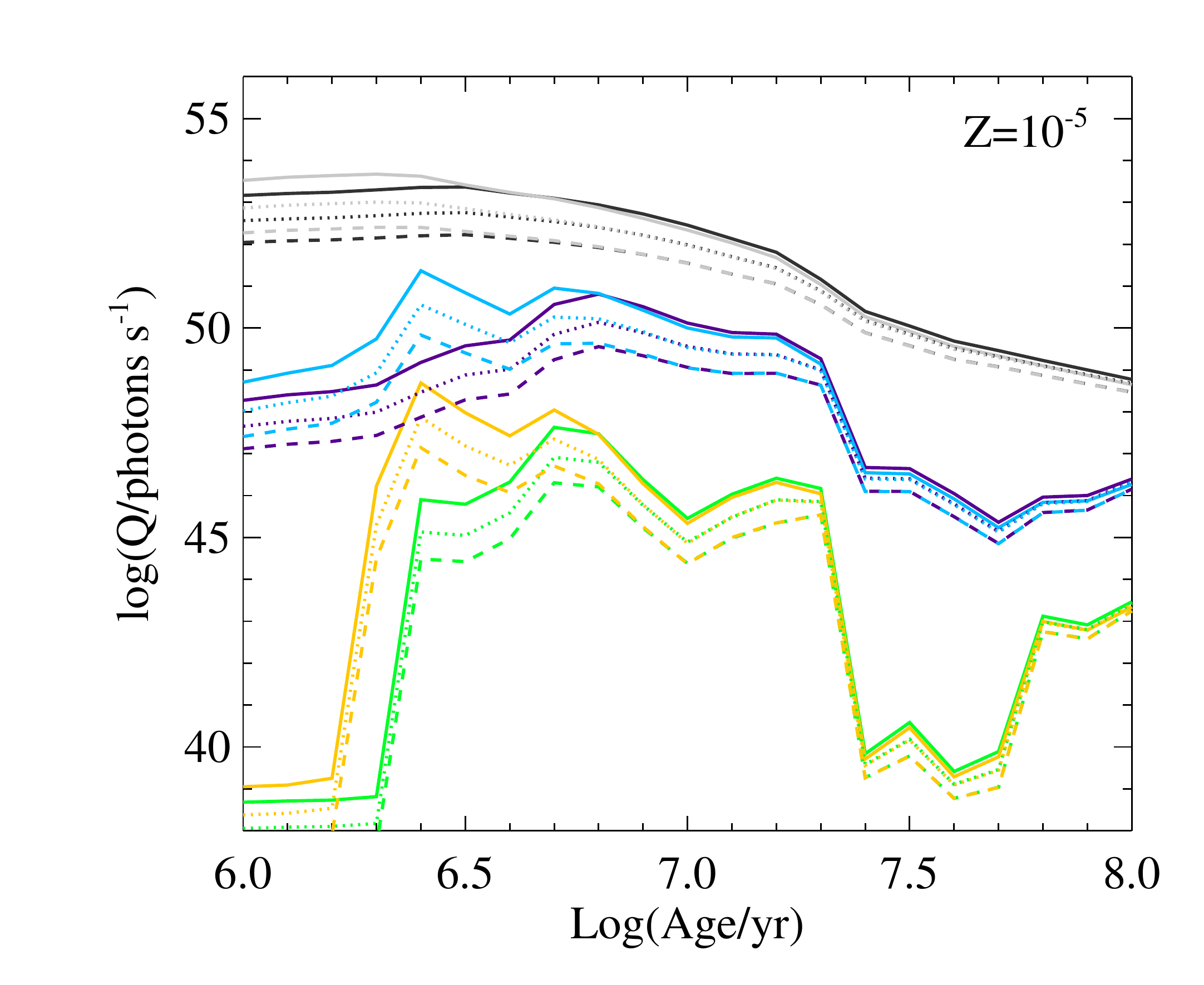}
    \includegraphics[width=0.45\textwidth]{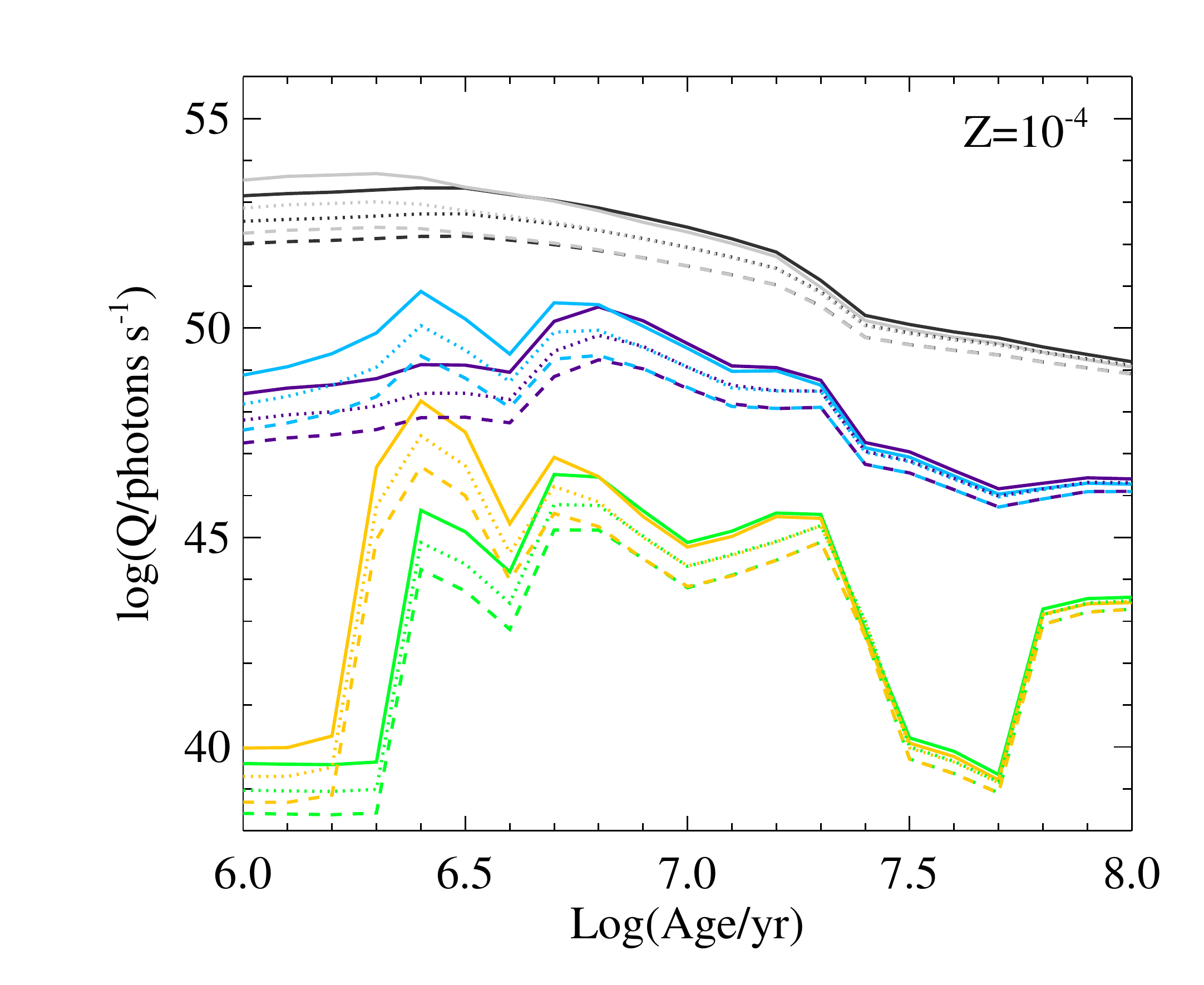}
    \includegraphics[width=0.45\textwidth]{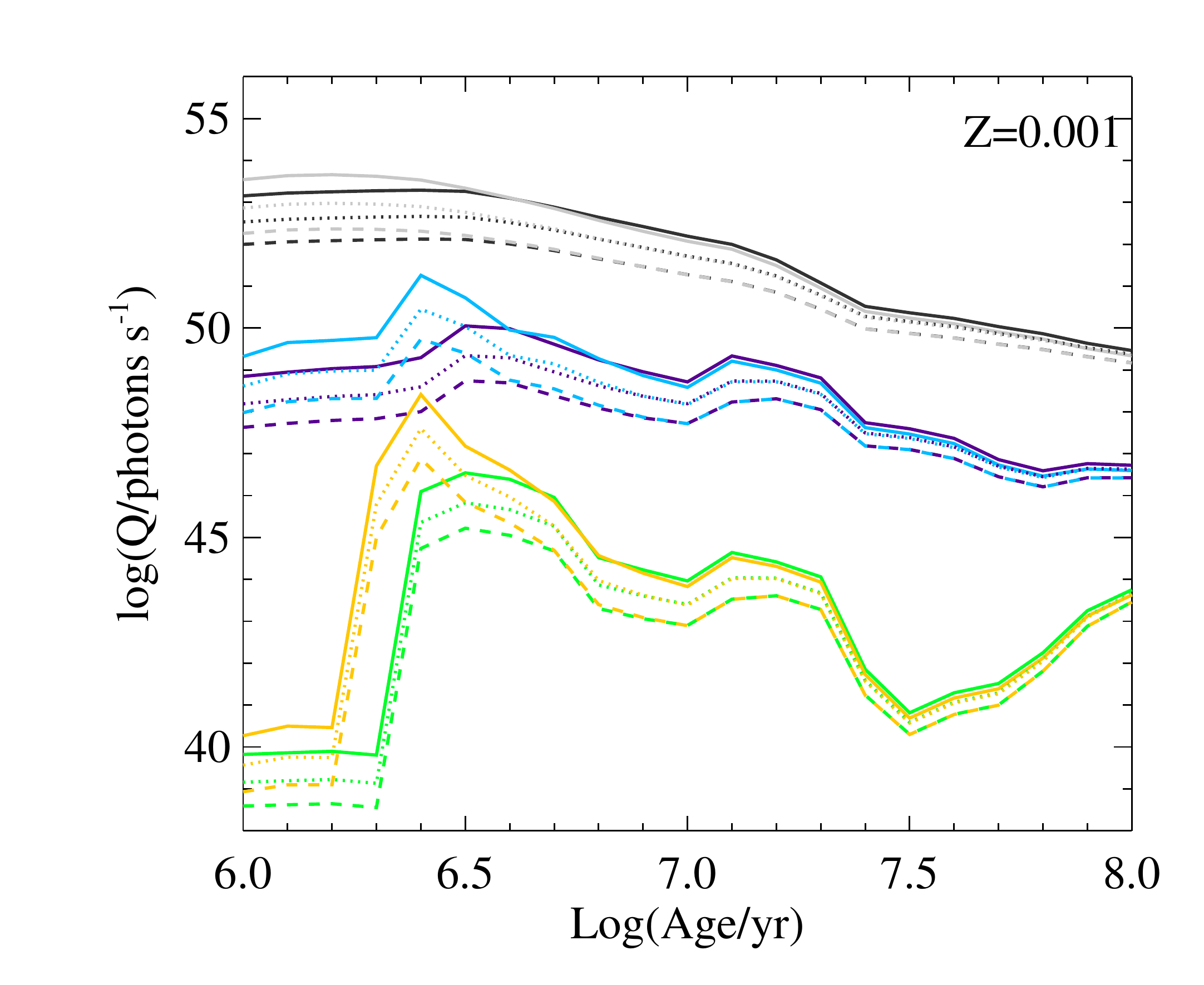}
    \includegraphics[width=0.45\textwidth]{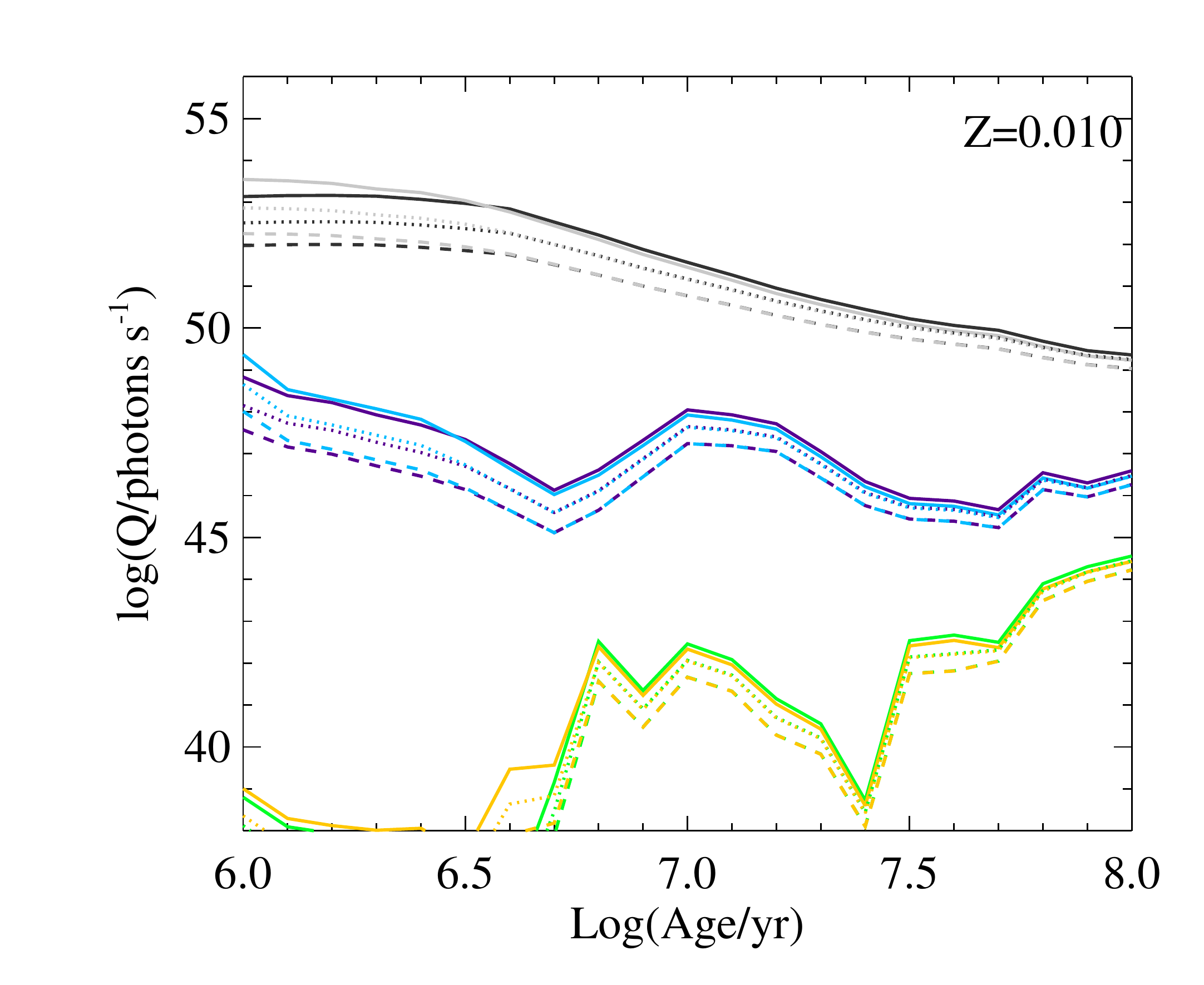}
    \includegraphics[width=0.45\textwidth]{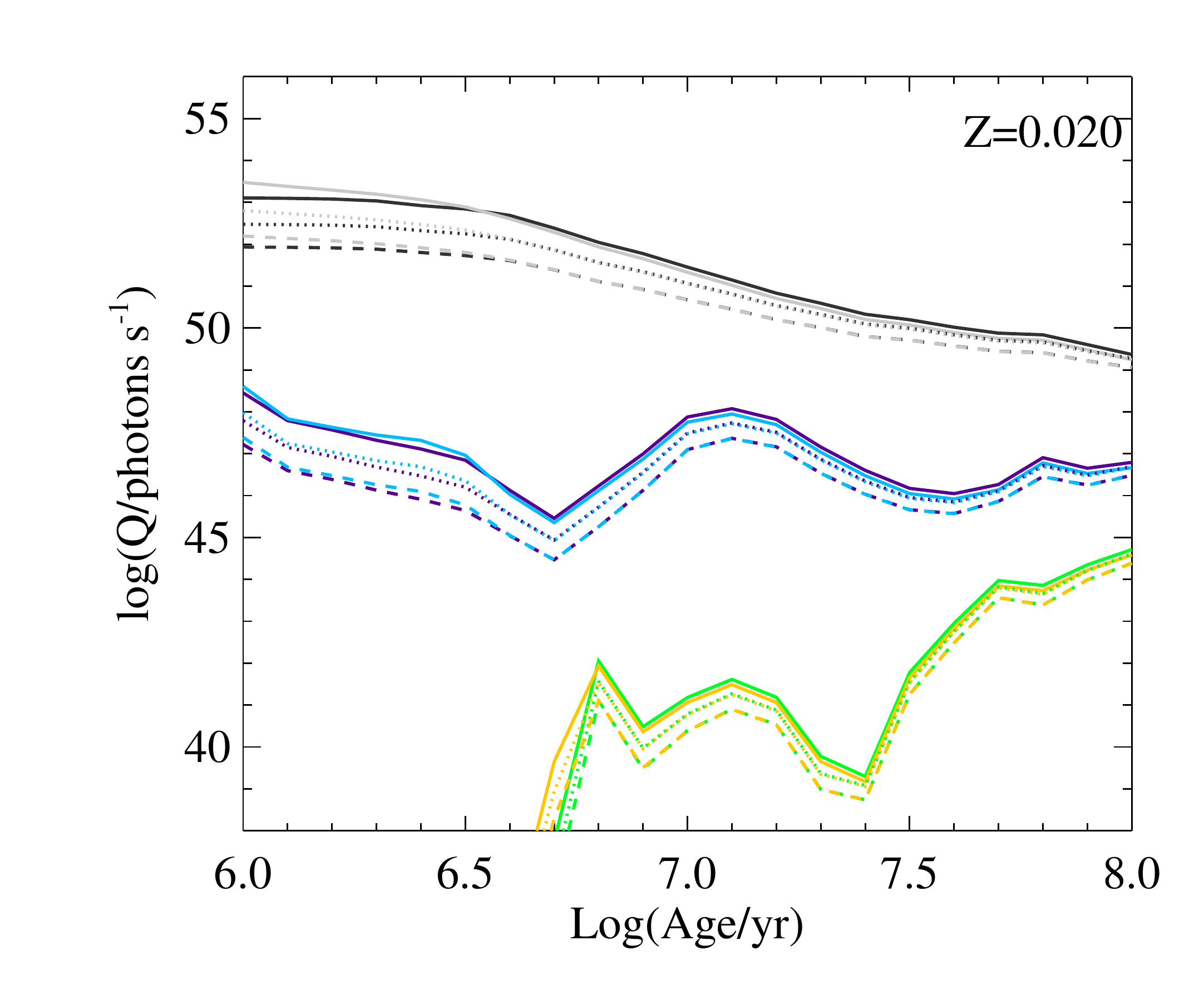}
    \includegraphics[width=0.45\textwidth]{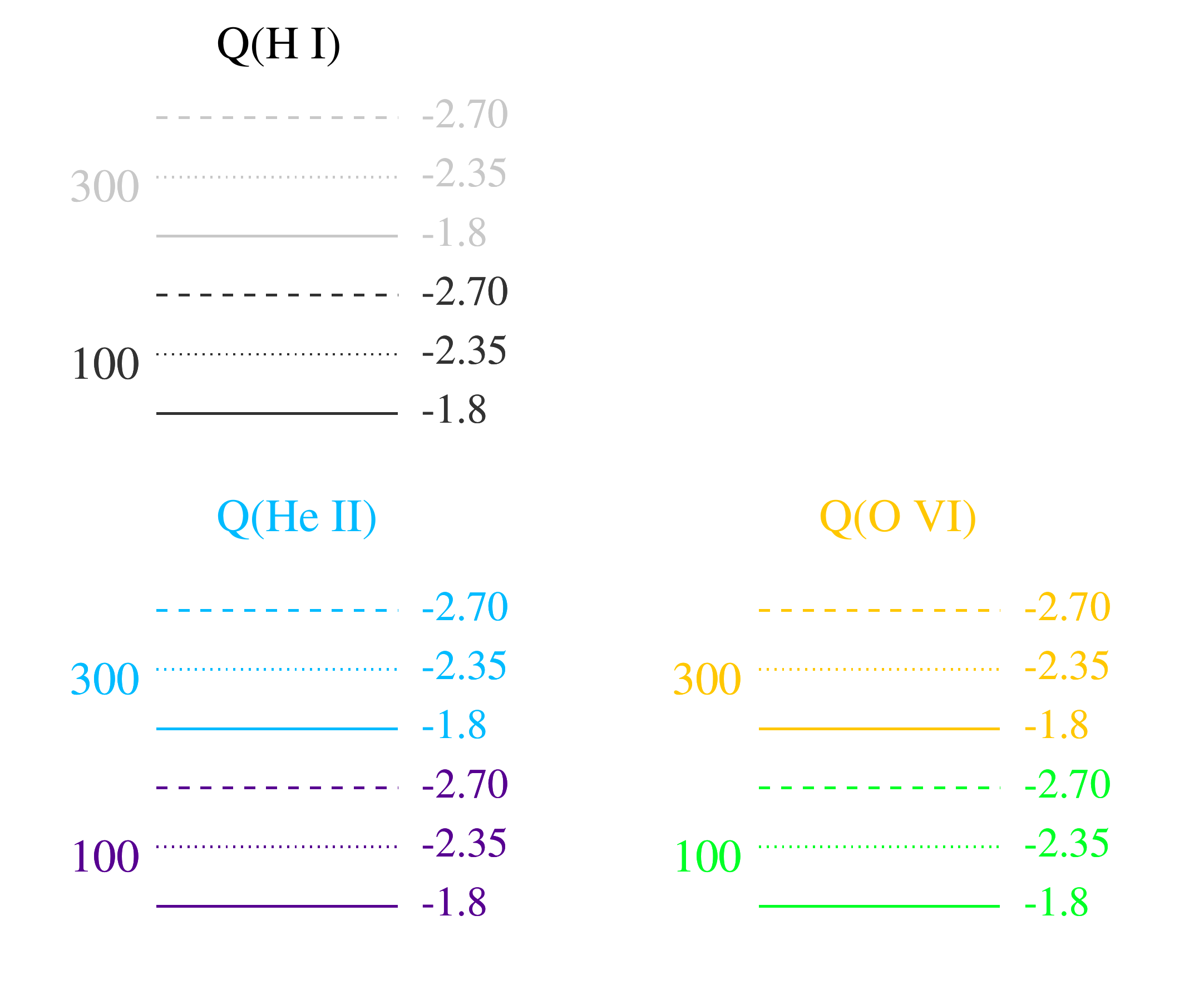}
 \caption{The time evolution of photon production rates in H I (black), He II (blue) and O VI (yellow-green) at five metallicities, from an instantaneous burst of total mass $10^6$\,M$_\odot$ occurring at an age of zero. At each metallicity three single-break power law ($\alpha_u=\alpha_m=-1.8, -2.35, -2.70$) IMF models are shown at two upper mass limits ($M_u=100, 300$\,M$_\odot$), with line style indicating different slopes and line colour different upper mass cut-offs as indicated in the key. The rest of the models lie within the bounds of those shown at fixed age.}\label{fig:time_evol}
 \end{figure*}

As shown in table \ref{tab:input_imfs}, we expand the standard BPASS IMF grid with power law models with single breaks at 0.5\,M$_\odot$ and $\alpha_m=\alpha_u$ in the range -1.8 to -2.7 at intervals of 0.1. We calculate each of these with four possible upper mass cut-offs: 100, 150, 200 and 300\,M$_\odot$, and for each of the output spectra calculate the time evolution of the ionizing photon flux from an instantaneous burst. As figure \ref{fig:time_evol} demonstrates, Q(H\,I) is largely independent of assumed IMF upper mass limit, and mildly sensitive to the upper mass slope, which changes the relative fraction of low mass stars with respect to their massive siblings.

By contrast, Q(He\,II) shows sensitivity to $M_u$, particularly at low metallcity and log(age/years)=6.3-6.5.  At these ages very massive stars are approaching the end of their life, swelling to become very luminous giants and (when stripped by winds or binary interactions) Wolf-Rayet stars. The presence or absence of these stars is strongly dependent on the upper mass cut-off.

The photon flux shortwards of 89.8\AA\ (138\,eV) capable of powering O\,VI emission and absorption, Q(O\,VI), is also sensitive to the presence or absence of these massive stars, both in terms of the peak flux attained by the population and its timing,as well as to metallicity. Stellar populations including very massive stars produce higher Q(O\,VI) and at substantially later ages than those which lack very massive stars. At high metallicities the strong stellar winds exhibited by very massive stars prevent them from ever becoming sufficiently hot or luminous to produce a significant Q(O\,VI) flux.

\begin{figure}
    %\centering
    \includegraphics[width=0.49\textwidth]{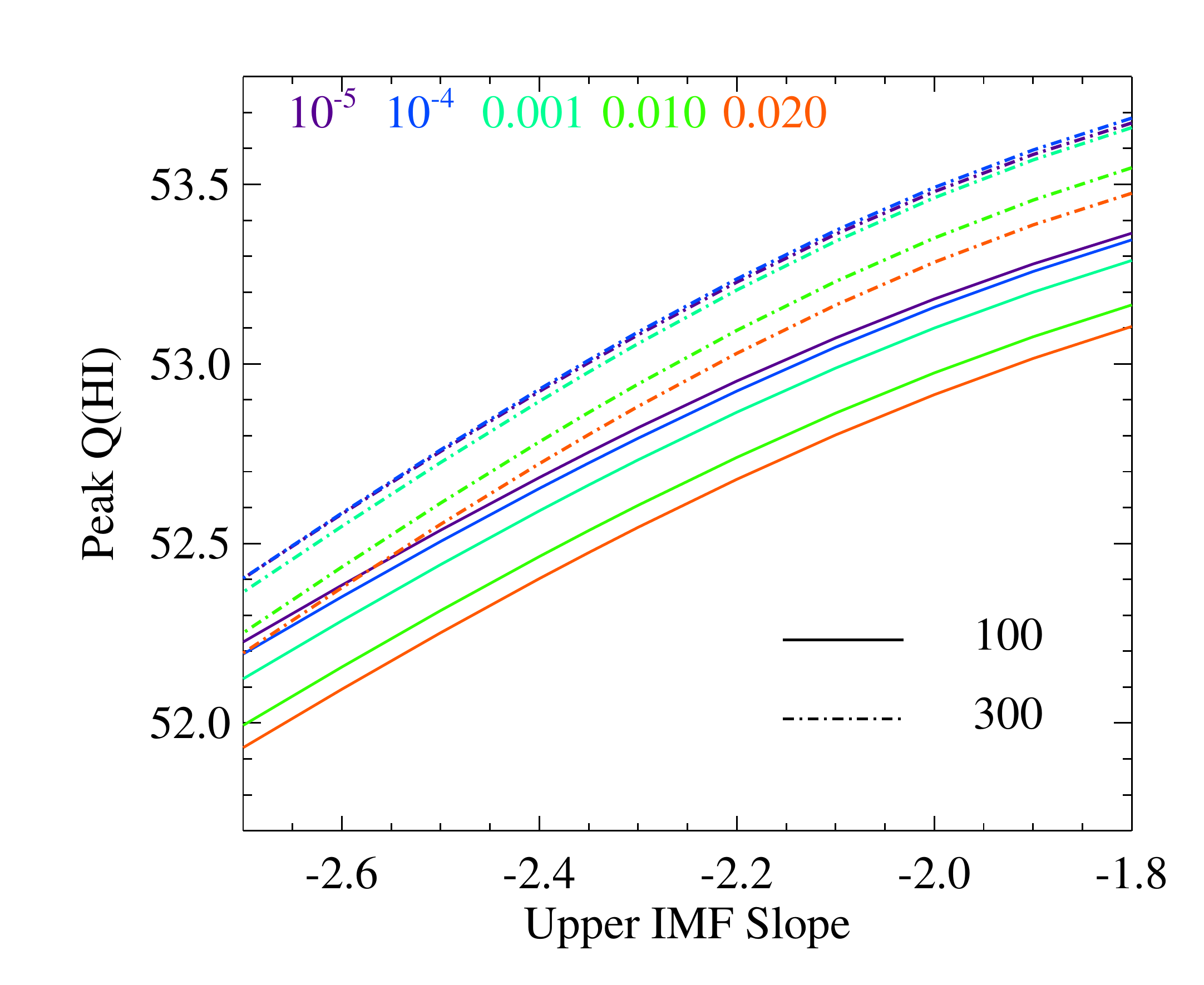}
    \includegraphics[width=0.49\textwidth]{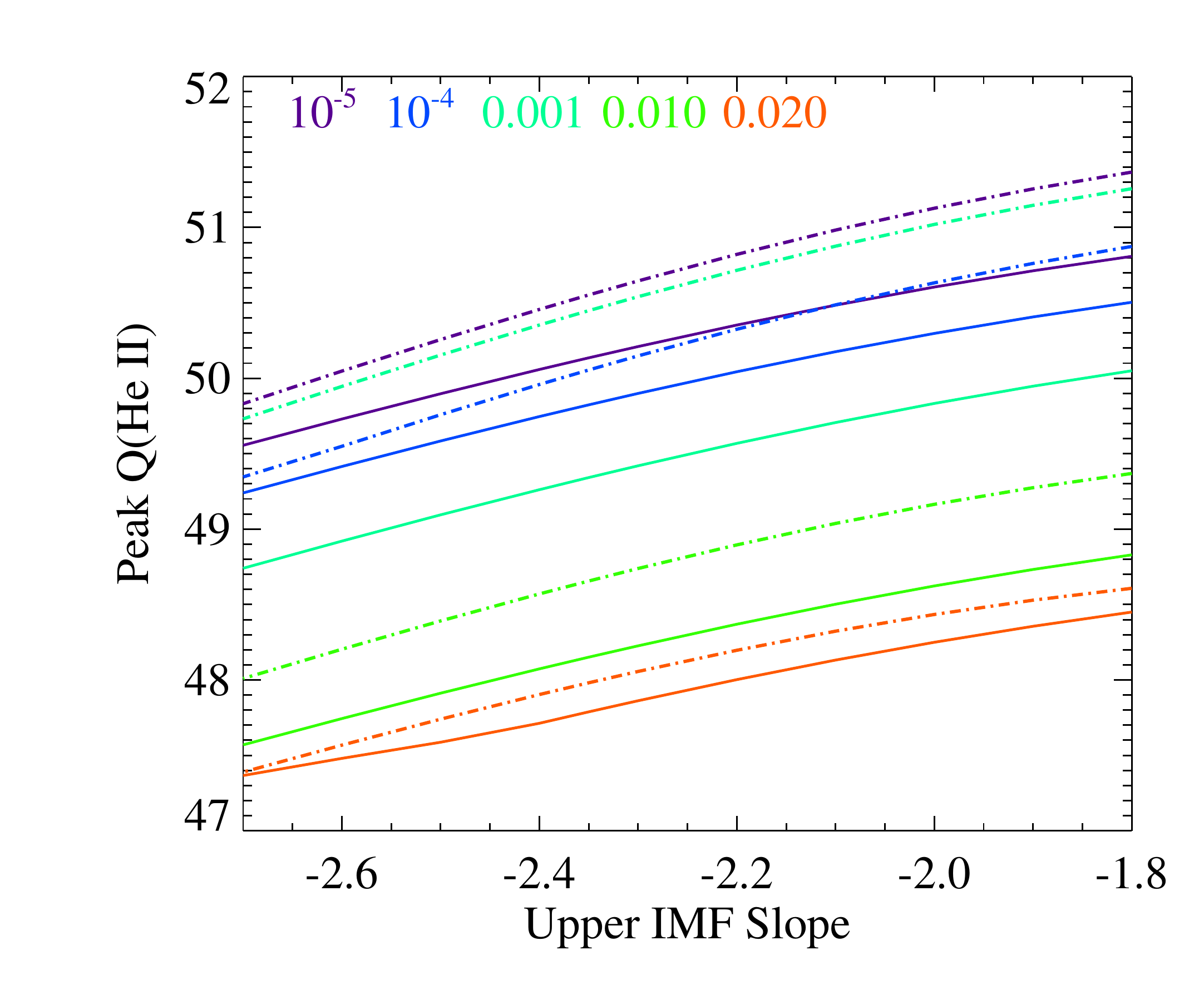}
    \includegraphics[width=0.49\textwidth]{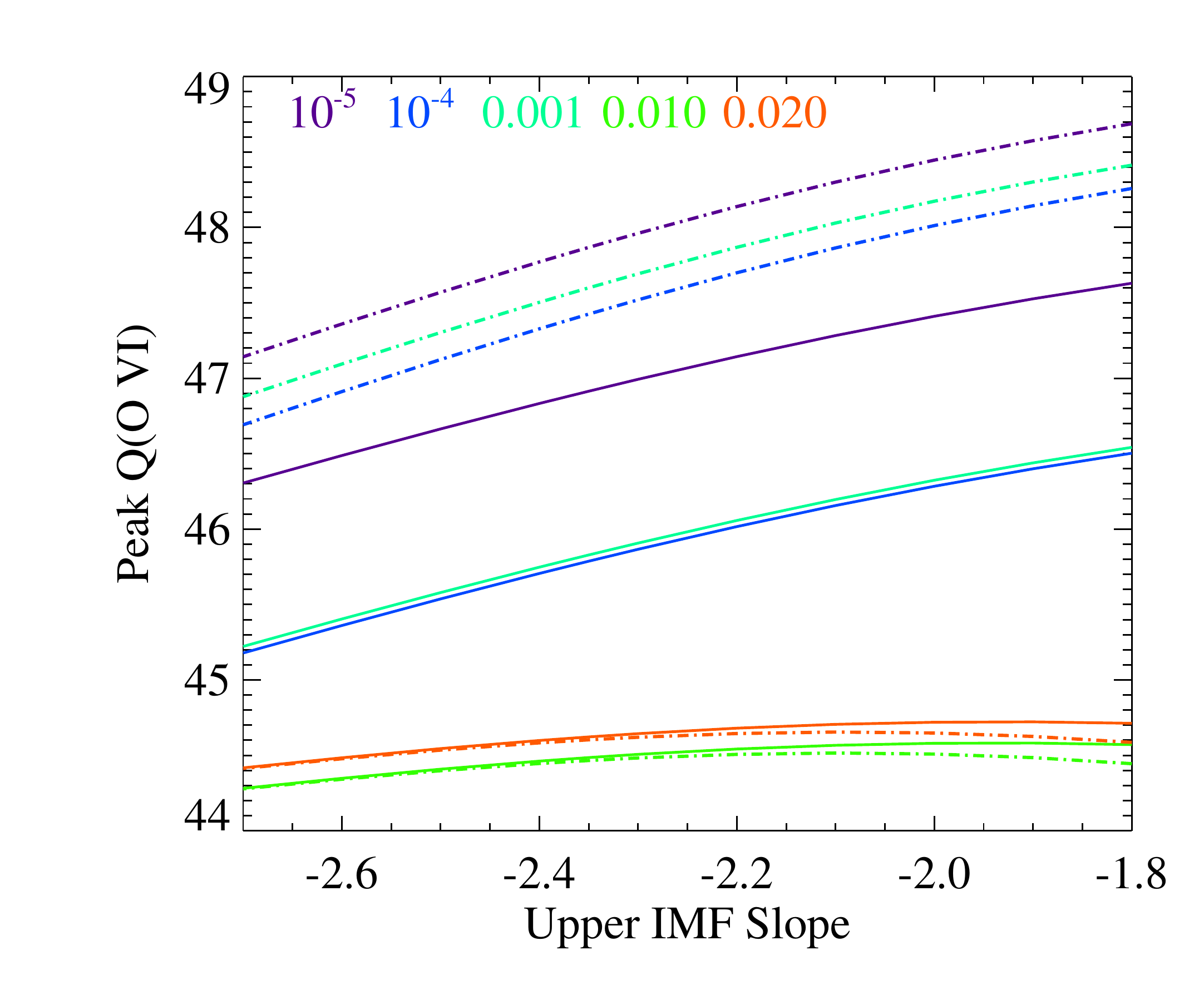}
 \caption{The peak ionizing photon fluxes reached by an ageing instantaneous starburst of total mass $10^6$\,M$_\odot$ as a function of IMF slope (assuming $\alpha_m=\alpha_u$). In each panel, metallicity is indicated by colour while upper mass cut-off $M_u$ is indicated by line style (100\,M$_\odot$ = solid, 300\,M$_\odot$ = dot-dash).}\label{fig:peak_fluxes}
 \end{figure}

In figure \ref{fig:peak_fluxes} we consider in more detail the effect of varying metallicity, IMF slope and upper mass limit in each photon flux. Here we show the peak flux obtained by the population in a given line, which is usually seen in a single time bin at log(age/years)=6.0-6.5, although we caution that the precise peak time may vary model by model. The strongest influence on hydrogen ionizing flux, Q(H\,I) appears to be the IMF slope, $\alpha_u$. Over the reasonable range of slopes explored here, peak Q(H\,I) varies smoothly by $\sim1.2-1.3$\,dex at each metallicity considered. By contrast, peak Q(H\,I) only varies by $\sim$0.3-0.4\,dex when the slope is fixed and the upper mass limit is altered. Overall, the range of peak fluxes is confined to a range of log(Q(H\,I))=51.9-53.7 for our $10^6$\,M$_\odot$ populations, narrower than that seen in Q(He\,II) and Q(O\,VI).

While the metallicity dependence of peak values of Q(He\,II) is perhaps the most visible feature of figure \ref{fig:peak_fluxes} (middle panel), this flux also shows more dependence on IMF slope ($\sim$1.2-1.6\,dex at fixed $M_u$) than upper mass cut-off ($\sim$0.6-1.2\,dex at fixed $\alpha_u$). The photon flux from the highest metallicity models (Z=0.020) is both very low and relatively stable to IMF variations. 

The IMF dependence of peak Q(O\,VI) flux is also governed by metallicity.  At the lowest metallicities considered here, IMF slope has a stronger effect than cut-off mass (1.6\,dex compared to 0.8\,dex). However at intermediate metallicities, around Z=0.001, the two effects are comparable ($\sim$1.6\,dex each). At the higher metallicities, $Z>0.010$, the ionizing flux is both very low and not dependent on IMF. In these cases, as figure \ref{fig:time_evol} shows, Q(O\,VI) is rising at late times and is generated by relatively low mass helium dwarfs, formed through binary interactions and largely independent of the very massive star population \citep[see e.g.][]{2018A&A...615A..78G}.

\begin{figure}
    \centering
    \includegraphics[width=0.49\textwidth]{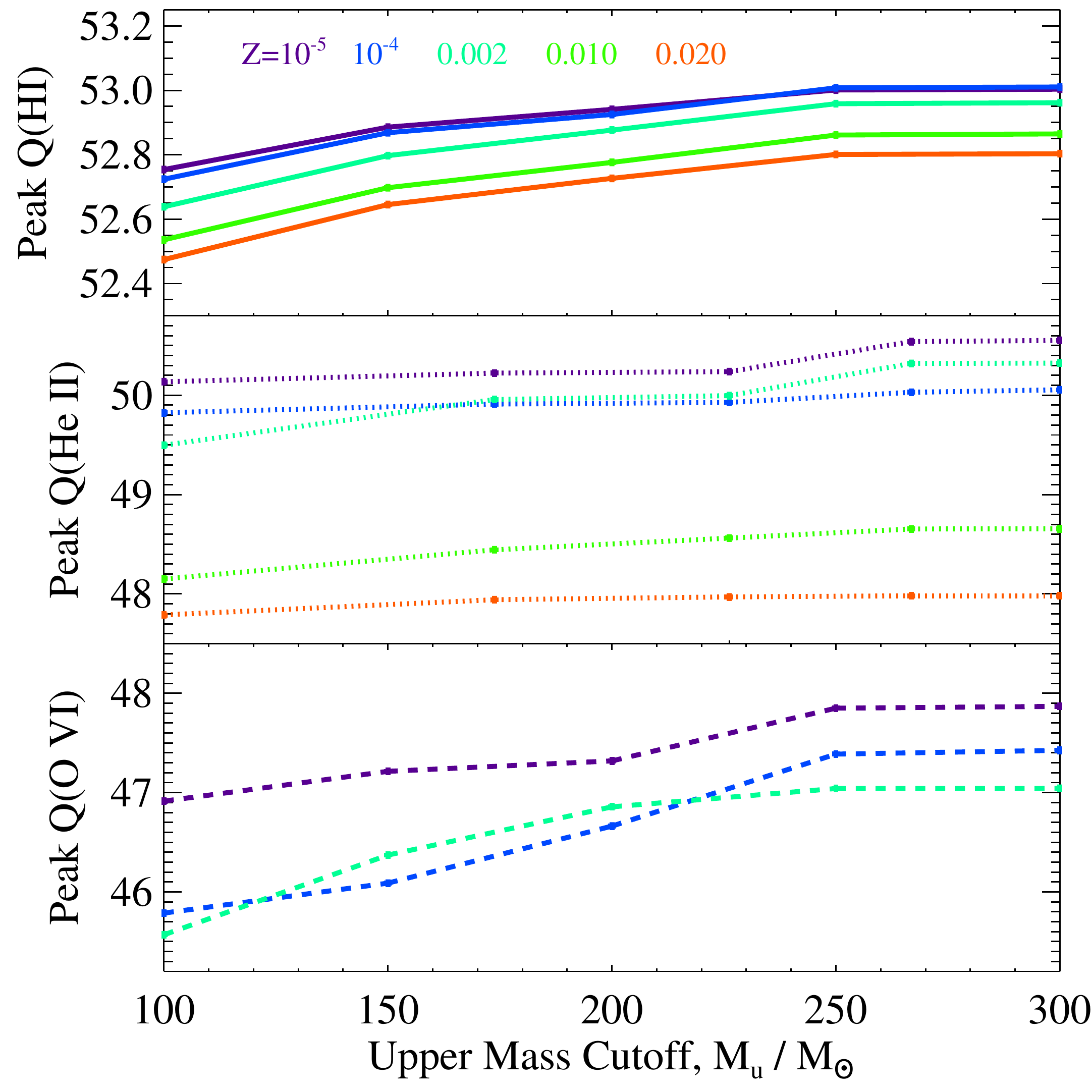}
 \caption{The peak ionizing fluxes, colour coded by metallicity, as in figure \ref{fig:peak_fluxes}, but now for single-break power law initial mass functions with a varying upper mass limit, $M_u$. We fix $\alpha_m=\alpha_l=-2.35$. The high metallicity fluxes in the He\,II and O\,VI ionizing photons are too low to show here, but follow a similar trend.}\label{fig:peak_mup}
 \end{figure}

While models with upper mass limits $M_u=100$ and $300$\,M$_\odot$ provide the bounding cases at fixed power law slope and metallicity, we also consider two intermediate upper mass cut-offs at  $M_u=150$ and $200$\,M$_\odot$. We show the variation in peak fluxes as a function of upper mass limit and metallicity for a fixed IMF slope ($\alpha_m=\alpha_u=-2.35$) in figure \ref{fig:peak_mup}. The peak H\,I ionizing flux increases smoothly with upper mass cut-off at all metallicities considered here. At most metallicities, the same is true of peak He\,II flux and O\,VI flux but anomalous behaviour emerges in our models with $Z\sim0.001$. At this metallicity (5\% of Solar) there is a very strong dependence of hard ionizing flux on upper mass limit which results from a combination of physical effects. These models have sufficient metal line opacities in their stellar atmospheres to drive mass loss through stellar winds, but at rates sufficiently low to allow angular momentum to be retained and rotational mixing to strongly influence the evolution of stars. In BPASS models, this chemically homogenous evolution is deemed to occur if significant mass is accreted by stars with $Z<0.004$ and $M>20$\,M$_\odot$. The result is that where the upper mass cut-off is as low as 100\,M$_\odot$ the majority of massive stars quickly drop below the BPASS mass limit for onset of chemically homogenous evolution and very few rotationally mixed stars are formed in our models. By contrast, when $M_u=300$\,M$_\odot$ a reasonable amount of mass loss can occur through winds, while still leaving stars with sufficient mass for rotational mixing to influence their evolution. Since chemically homogenous stars tend to burn hotter and for longer than a non-rotating star of comparable mass, these stars dominate emission at wavelengths capable of ionizing He\,II. The result is a strong dependence on $M_u$ in the hard ionizing flux at $Z=0.001$ relative to those at higher metallicities (where stellar winds always dominate and fewer stars are stripped) and those at the lowest metallicities.
At $Z\leq10^{-4}$, stellar winds are very weak due to the low metal ion opacities in the stellar atmospheres, and very little of the initial stellar mass is lost to winds. These stars are more likely than those at higher metallicities to be rotationally mixed, but those at the lowest metallicity also tend to become inflated to larger stellar radii which increases the probability that their atmospheres may become stripped by a binary companion (and so again create a hot, blue spectrum). The detailed balance of these different physical mechanisms for producing hot stars leads to a somewhat counterintuitive inversion in the metallicity trends of Q(He\,II) and Q(O\,VI) with the $Z=0.001$ models producing more hard ionizing photons than those at $Z=10^{-4}$.

%________________________________________________________________

\section{Introducing a Second Break}\label{sec:doublebreak}

Studies of high mass star forming regions in the local Universe, notably the 30 Doradus complex in the Large Magellanic Cloud, have suggested that the initial mass function for massive stars may be substantially shallower than the Salpeter-like initial mass function that appears common for near-Solar mass stars \citep{2018arXiv180703821S,2018Sci...359...69S}. A possible interpretation of these findings is that the power-law initial mass function may show a second break in the high- to very-high-mass regime.

\begin{figure}
    \centering
    \includegraphics[width=0.49\textwidth]{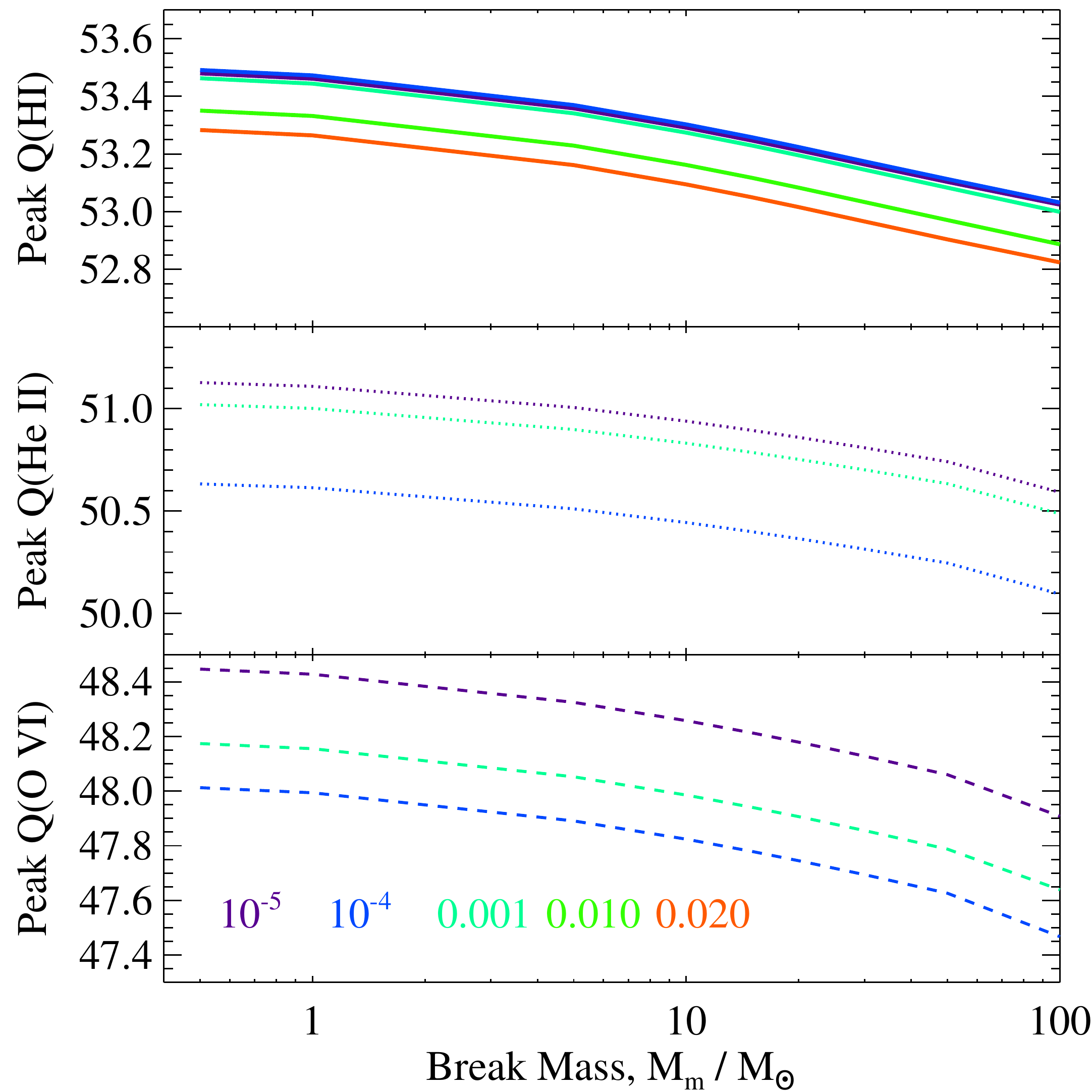}
 \caption{The peak ionizing fluxes, as in figure \ref{fig:peak_fluxes}, but now for power law initial mass functions with two breaks. We fix $M_0=0.5$\,M$_\odot$, $M_u$=300\,M$_\odot$, $\alpha_l=-1.3$, $\alpha_m=-2.35$ and $\alpha_u=-2.0$, while allowing the intermediate break mass, $M_m$, to vary. The high metallicity fluxes in the He\,II and O\,VI ionizing photons are too low to show here, but follow the same trend.}\label{fig:peak_fluxes2}
 \end{figure}

To explore the effects of such an IMF, we calculate a limited set of models with two power law breaks.  We fix $M_0=0.5$\,M$_\odot$, $M_u$=300\,M$_\odot$, $\alpha_l=-1.3$, $\alpha_m=-2.35$ and $\alpha_u=-2.0$ for consistency with constraints from local studies, while allowing the intermediate break mass, $M_m$, to vary. Again, we look at the peak photon fluxes attained by the population as a function of metallicity, and break mass, as shown in figure \ref{fig:peak_fluxes2}.

Unsurprisingly, the effects of introducing a second break on peak photon flux are relatively minor. The model fluxes lie between those of single-break power law initial mass functions with slopes of $\alpha_m=\alpha_u=-2.0$ and $-2.35$, tending towards one or the other depending on the intermediate break mass, $M_m$ selected. There is no clear sign of distinctive behaviour of double-break models at the metallicities considered, and in the absence of other probes these would likely be interpreted as single break models with a slope between the two limits if the appropriate photon flux ratios were observed.

%________________________________________________________________

\section{Effects of Total Stellar Mass}\label{sec:mmax}

BPASS, in common with most other stellar population synthesis models, assumes that the star forming population is sufficiently massive that the IMF is fully sampled. An obvious problem with this assumption arises in the case of small starbursts; if the total mass of a newborn stellar population is only 1000\,M$_\odot$, the probability of forming a single 300\,M$_\odot$ star is very low. The fully sampled statistical IMF approach is equivalent to assigning fractions of the flux of such very massive stars to the composite spectrum, rather than considering a binary probability (1 or 0) that the star contributes. Since even a single individual massive star can dominate the hard ionizing radiation from a population, and the spectrum of that star is strongly mass dependent, this presents a challenge to modelling.

An alternate approach is stochastic sampling - drawing individual stars from an underlying probability distribution until the desired total mass is reached \citep{2013NewAR..57..123C}. Because this is fundamentally probabilistic, many iterations must be performed for each initial mass function and each total mass under consideration, in order to form a complete picture. While models which do this exist and have been used to explore the properties of the ultraviolet continuum \citep[e.g.][]{2003A&A...407..177C,2009MNRAS.395..394P,2010A&A...517A..93V,2011ApJ...741L..26F,2012ApJ...750...60F,2017MNRAS.470.3532V,2017MNRAS.470.1612P,2018MNRAS.480.3091A}, these have been limited to treatment of single star populations.  A detailed stellar evolution-based code such as BPASS is ill-suited for such a procedure.  Reading each stellar model in turn, matching it with an atmosphere model at each timestep and calculating its contribution to the composite spectrum in a given time bin is a slow process which does not lend itself to many iterations or large grids of parameters. Experimentation with an earlier BPASS version suggested that the effects of binary interactions broaden the range of evolutionary pathways at each initial mass and allow for stars to become more massive via mass transfer and mergers that mitigate some of the stochastic sampling effects \citep{2012MNRAS.422..794E}.

Here we consider an intermediate approach between simple IMF weighting and full stochastic sampling. For a selected metallicity (Z=0.001), upper IMF slope ($\alpha_m=\alpha_u=-2.35$) and nominal mass cut-off ($M_u$=300\,M$_\odot$), we consider hybrid initial mass functions in which the weighting distribution for individual models is generated assuming full sampling but then modified. Given a range of assumed total stellar masses ranging from 100 to $10^8$\,M$_\odot$, we remove individual models from the spectral synthesis if their contribution to a stellar population of that initial mass is less than 0.5 stars. We then readjust the overall scaling to account for the diminished mass of stars contributing, giving the final results as the peak photon flux per stellar mass formed in the initial starburst.

In the case of a single star population, in which the probability of a model occurring only depends on the stellar initial mass, this approach is equivalent to an upper mass cut-off, $M_u$, that scales with the total starburst size. As a binary population synthesis code, BPASS includes a large number of models at each initial mass, with weightings which also scale with the (primary mass-dependent) period and mass ratio distributions. As a result, while the highest mass stars will always be excluded from a small population, some intermediate mass primaries will be represented by a subset of period and mass ratio models, while less frequently occurring models will be omitted.

We note that this analysis does not account for the full stochastic scatter in the population. While the probability of each high mass stellar model occurring in a small population may be low, there are likely to be a few individual stars in the high mass range which this prescription will not capture. In rare cases, a low mass population will host a large number of massive stars; in the majority of cases it will not. Similarly it is possible for a high mass starburst to lack any high mass stars when stochastic sampling is taken into acount. Thus the models calculated here are indicative of the scale of the uncertainties rather than their spread.

\begin{figure}
     \centering
    \includegraphics[width=0.49\textwidth]{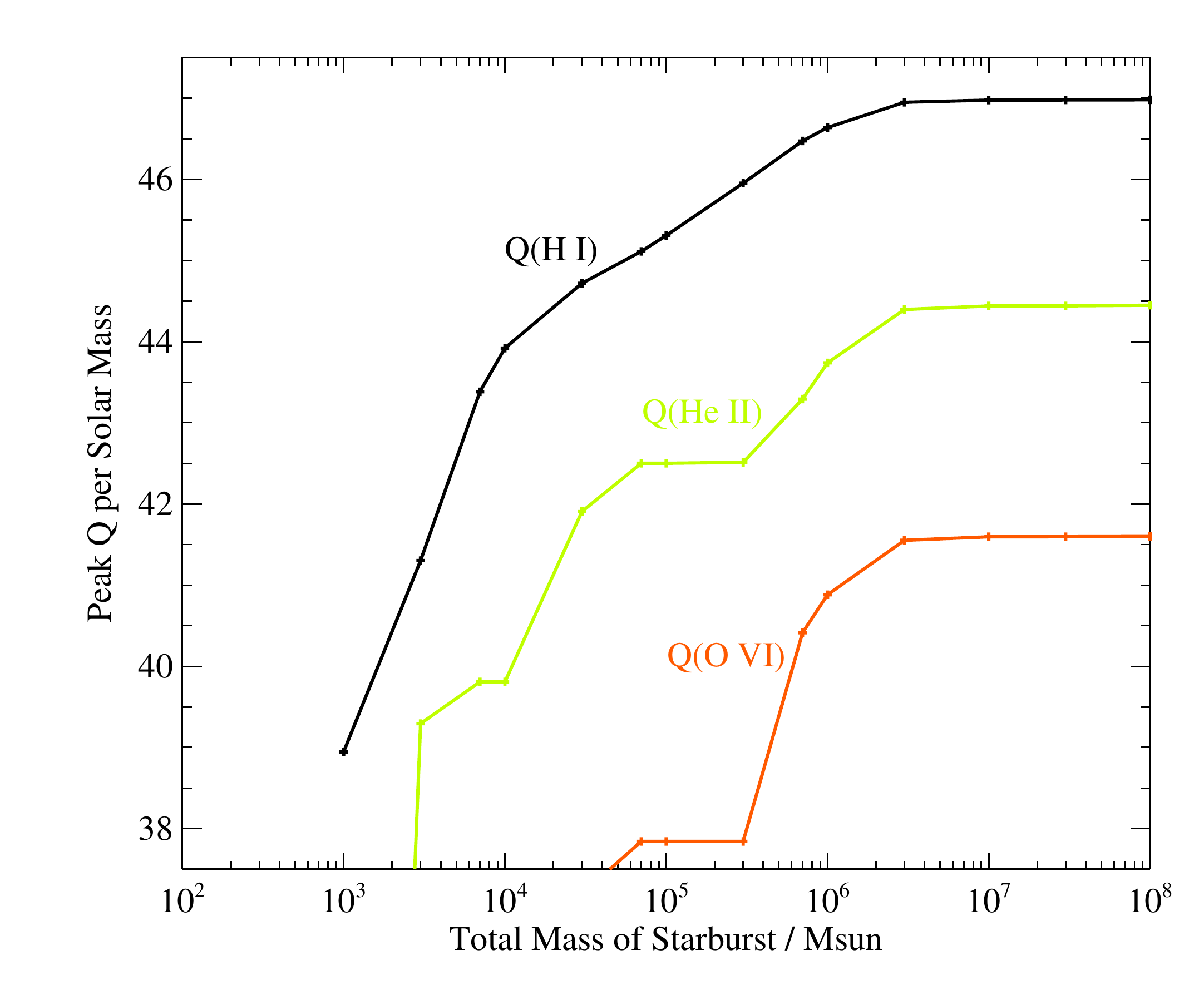}
  \caption{The peak ionizing photon fluxes per Solar mass of stars, reached by an ageing instantaneous starburst at $Z=0.001$ as a function of initial total mass assuming a total-mass-dependent cut-off in the high mass stellar population as described in section \ref{sec:mmax}.}\label{fig:mmax}
 \end{figure}

The results of this analysis are shown in figure \ref{fig:mmax}. As the figure indicates, the ionizing flux from a population does not become a simple product of total stellar mass until $M_\mathrm{total}\ge10^6$\,M$_\odot$, at which point the IMF is fully populated. The Q(He\,II) and Q(O\,VI) flux from BPASS models show plateaus at intermediate masses. This is likely a modelling artifact resulting from the relatively coarse grid of BPASS stellar models at high masses, and emphasizing the importance of individual very massive stars in producing these lines. The transition between the ionizing photon flux from a population hosting no very massive stars to one hosting even a single 300\,M$_\odot$ star is abrupt and dramatic in this implementation. As a result, it seems likely that stochastic sampling of the initial mass function may be important for interpretation of high ionization potential spectral features, even in the galaxy-wide starbursts seen in the distant Universe. This has been seen before in work exploring spectral synthesis with single star evolution models \citep[e.g.][]{2017MNRAS.470.3532V,2017MNRAS.470.1612P}  but the analysis here suggests that the result holds true for binary models, despite the increased range of evolutionary pathways: the ionizing flux of the most massive star is simply too dominant in contributing the hard ionizing photons under consideration.

\begin{figure*}
    \centering
    \includegraphics[width=0.32\textwidth]{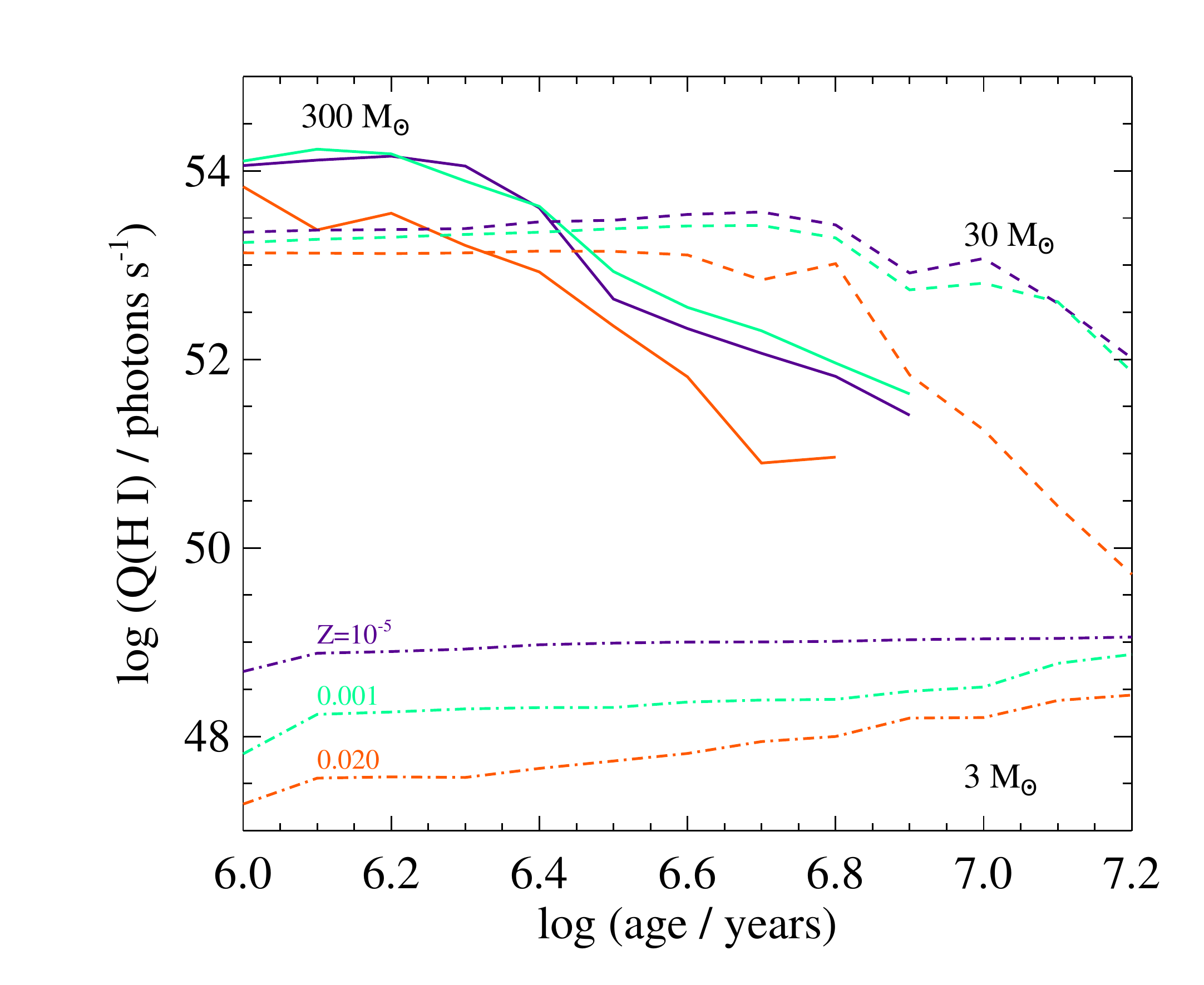}
    \includegraphics[width=0.32\textwidth]{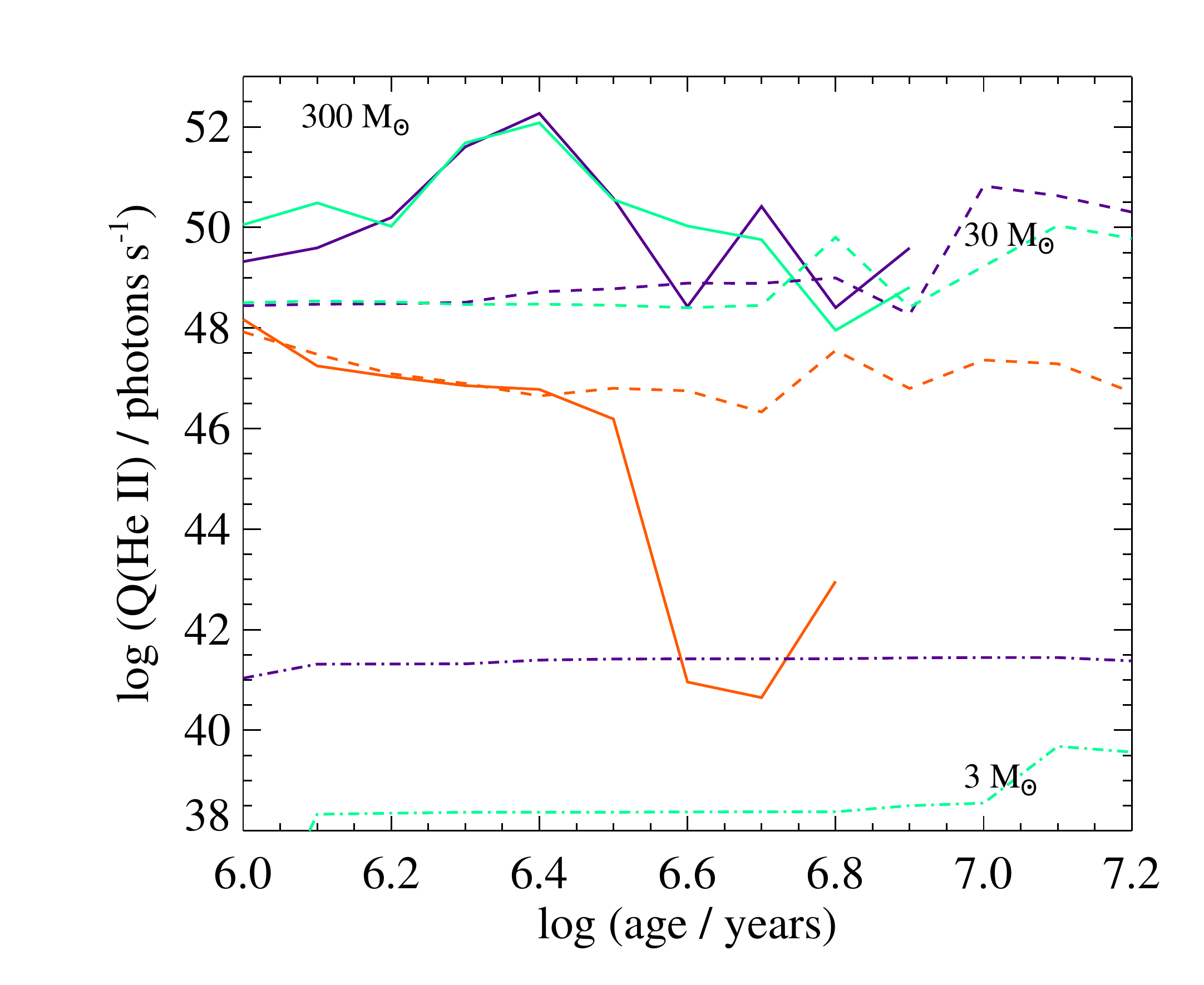}
    \includegraphics[width=0.32\textwidth]{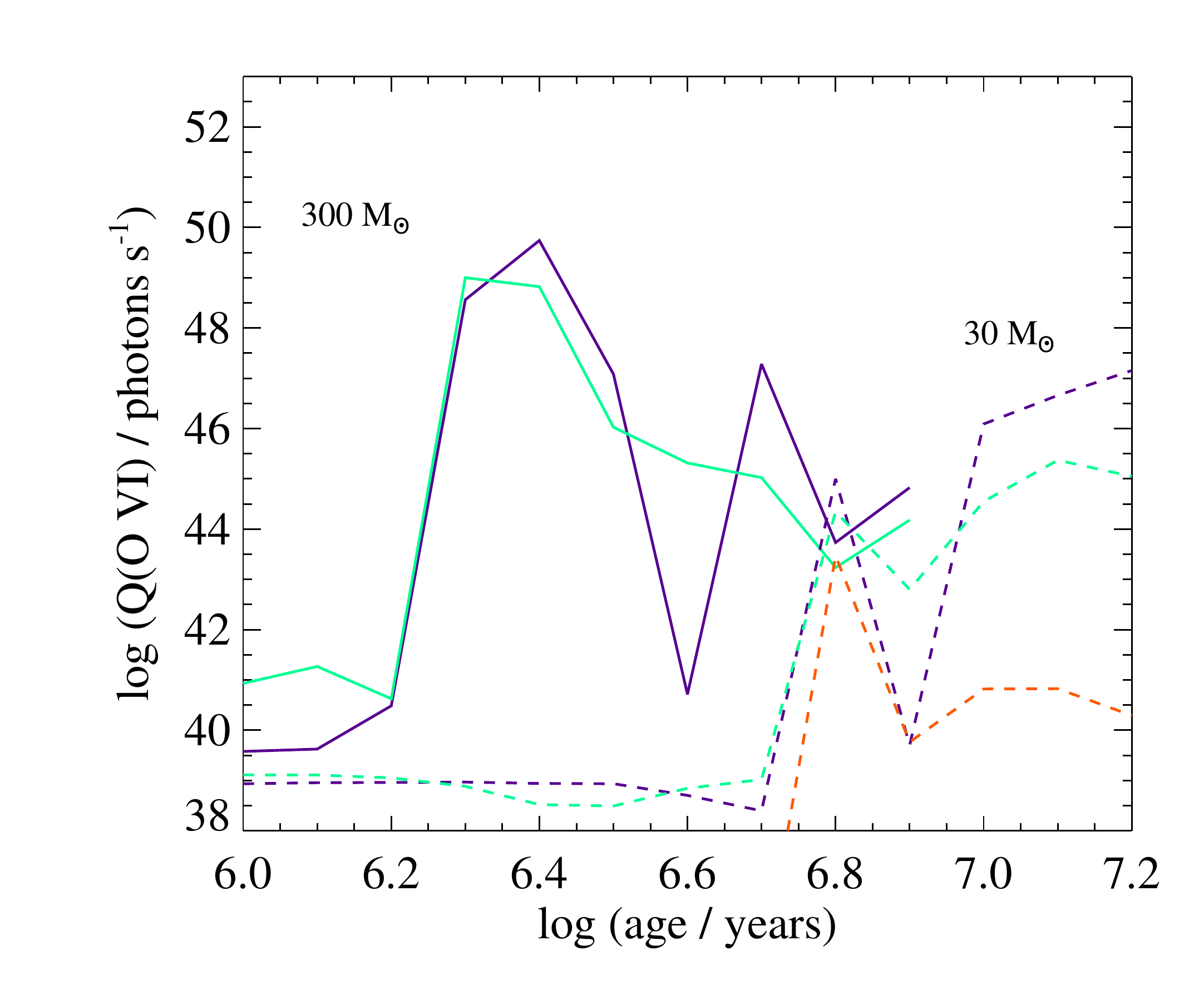}
  \caption{The ionizing photon flux time evolution for stellar populations comprised entirely of single stars or primaries at a given mass (3, 30 or 300\,M$_\odot$) and appropriate binary companions. Three metallicities are shown for each mass, and each energy threshold. For Q(O\,VI) the emission from low mass and high metallicity stars is negligible.}\label{fig:star}
 \end{figure*}

To test the limits of stochastic sampling, we generate three additional population synthesis models. In the first, only 300\,M$_\odot$ primary stars and their binary companions (given by the standard BPASS v2.2 binary parameter distribution) are included in a population of total mass $10^6$\,M$_\odot$, in the second, only 30\,M$_\odot$ stars (and their companions) and in the third, only 3\,M$_\odot$ stars (single and primary, with their binary companions). The ionizing fluxes from these three models as a function of metallicity are shown in figure \ref{fig:star}. The short lifetimes of the 300\,M$_\odot$ stars and their rapid evolution leads to extreme variation on their hard ionizing flux in short timescales, even while the Q(H\,I) hydrogen ionizing flux declines slowly and smoothly. Stars at 30\,M$_\odot$ have hydrogen ionizing fluxes almost an order of magnitude lower than the very massive stars, but this is sustained over a much longer lifetime, as is Q(He\,II). The only photons capable of ionizing O\,VI from this population arise from the stripped binary products present at relatively late times. Finally, the 3\,M$_\odot$ population produces very little flux shortwards of the Lyman limit, and negligible emission in Q(O\,VI). All stellar populations show much lower ionizing fluxes at Solar metallicity (Z=0.020) than at lower metallicities.
The lowest ratio in Q(H\,I)/Q(He\,II) attained by the 300\,M$_\odot$ population at $Z=10^{-5}$ is 1.3 dex.

%________________________________________________________________

\section{Discussion}\label{sec:disc}

\begin{figure}
    \centering
    \includegraphics[width=0.49\textwidth]{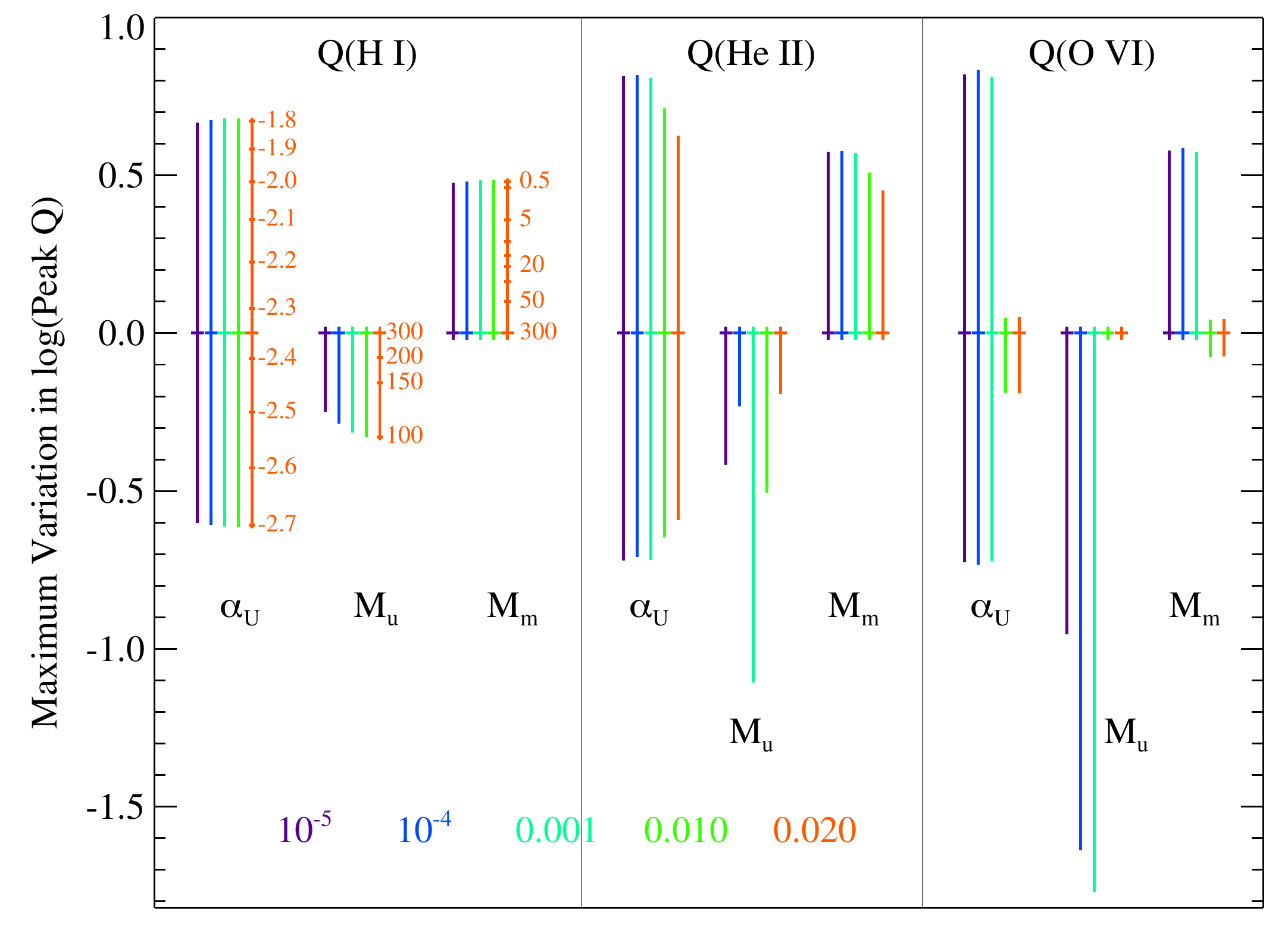}
  \caption{The range of variation in peak ionizing photon fluxes across the grid of models considered here. Range is shown separately for each metallicity, first for single-break IMFs with $M_u=300$\,M$_\odot$ and varying $\alpha_m=\alpha_l$, then for models with fixed $\alpha_m=\alpha_l=-2.35$ and varying $M_u$, and finally for double break models with $\alpha_m=-2.35$, $\alpha_l=-2.00$ and $M_u=300$\,M$_\odot$ but varying $M_m$. All ranges are shown relative to the BPASS default IMF model, which is a single break power law with $\alpha_m=\alpha_l=-2.35$ and $M_u=300$\,M$_\odot$.}\label{fig:var1}
 \end{figure}

The range of variation in peak fluxes, given our grid of `reasonable' initial mass function parameters is shown in figure \ref{fig:var1} at each metallicity, where the values are shown relative to the BPASS default IMF model, which is a power law with a single break at $M_l=0.5$\,M$_\odot$, $\alpha_m=\alpha_l=-2.35$ and $M_u=300$\,M$_\odot$. As the figure makes clear, the hydrogen ionizing flux $Q(H\,I)$ is relatively predictable in its response to IMF variations, as the effect of the IMF is almost independent of metallicity - in other words, if a value is calculated for one IMF and metallicity, the photon flux given different assumptions can be approximated by application of a simple scaling factor. By contrast, the hard ionizing flux is far more sensitive to variations in initial mass function and metallicity. In all cases, the default, Salpeter-like upper mass slope, model lies in the centre of the range of $\alpha_l$ variations, but our default cut-off $M_u=300$\,M$_\odot$ at the top end of the upper mass limit variations. At moderate metallicities ($Z=10^{-4}-0.010$), it is the upper mass cut-off that has the strongest effect on hard ionizing flux; very massive stars with M$>100$\,M$_\odot$ are critical for the production of photons sufficiently strong to power He\,II or O\,VI spectral features. The introduction of a second break to a shallower slope at high masses can also boost hard photon production by the same mechanism: increasing the number of very massive stars in the population.

\begin{figure}
    \centering
    \includegraphics[width=0.49\textwidth]{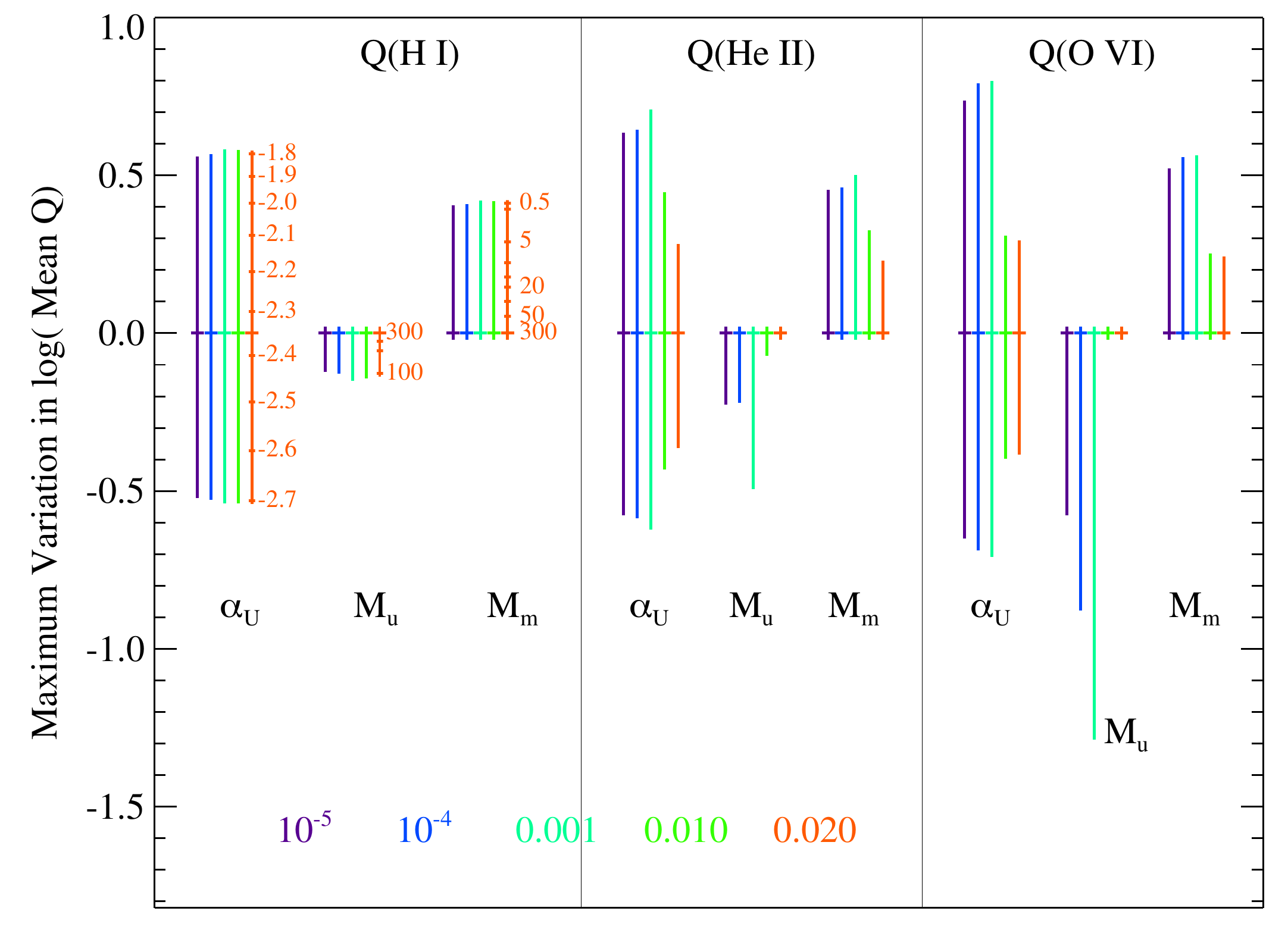}
  \caption{As in figure \ref{fig:var1}, but now considering the mean photon flux averaged over the first 20\,Myr of the starbursts' evolution.}\label{fig:var2}
 \end{figure}

While the peak fluxes attained by a population may be critical for interpreting some of the hardest observed spectra in both the distant and local Universe, for the majority of star forming regions, this is attained only for a very brief interval in the starbursts' evolution. A more generally useful quantity may be the time-averaged photon flux over the first 20\,Myr of stellar evolution (i.e. at log(age/years)$<$7.3). As figure \ref{fig:time_evol} demonstrated, this interval is sufficient to capture the lifetimes of the most massive stars, and hence the interval in which hard fluxes might be expected.  In figure \ref{fig:var2} we show the equivalent variation in this quantity. As might be expected, it shows very similar trends to those seen in the peak fluxes, but with the variation mitigated by the time average, yielding somewhat narrower ranges.

\begin{figure*}
    \centering
    \includegraphics[width=0.45\textwidth]{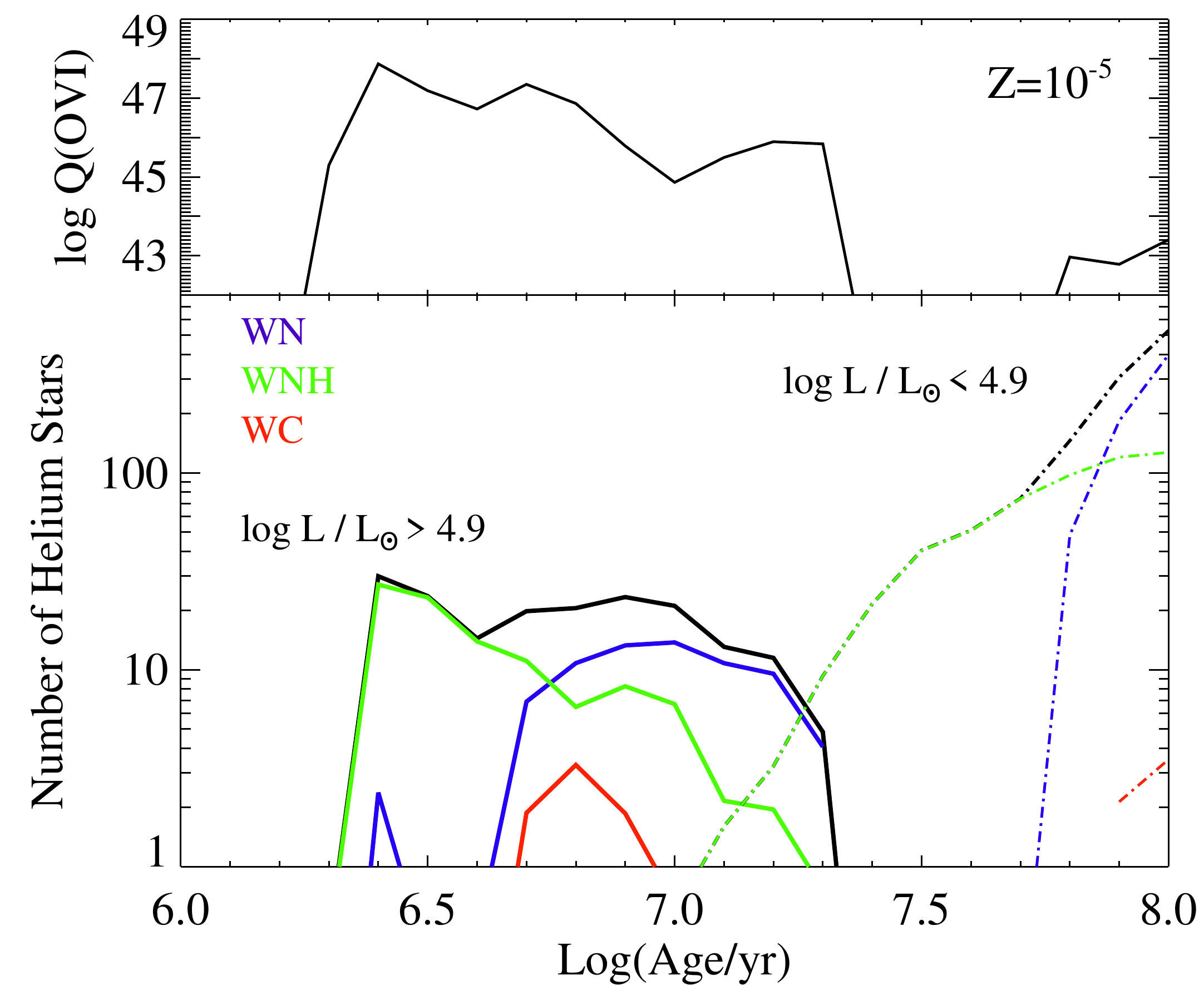}
    \includegraphics[width=0.45\textwidth]{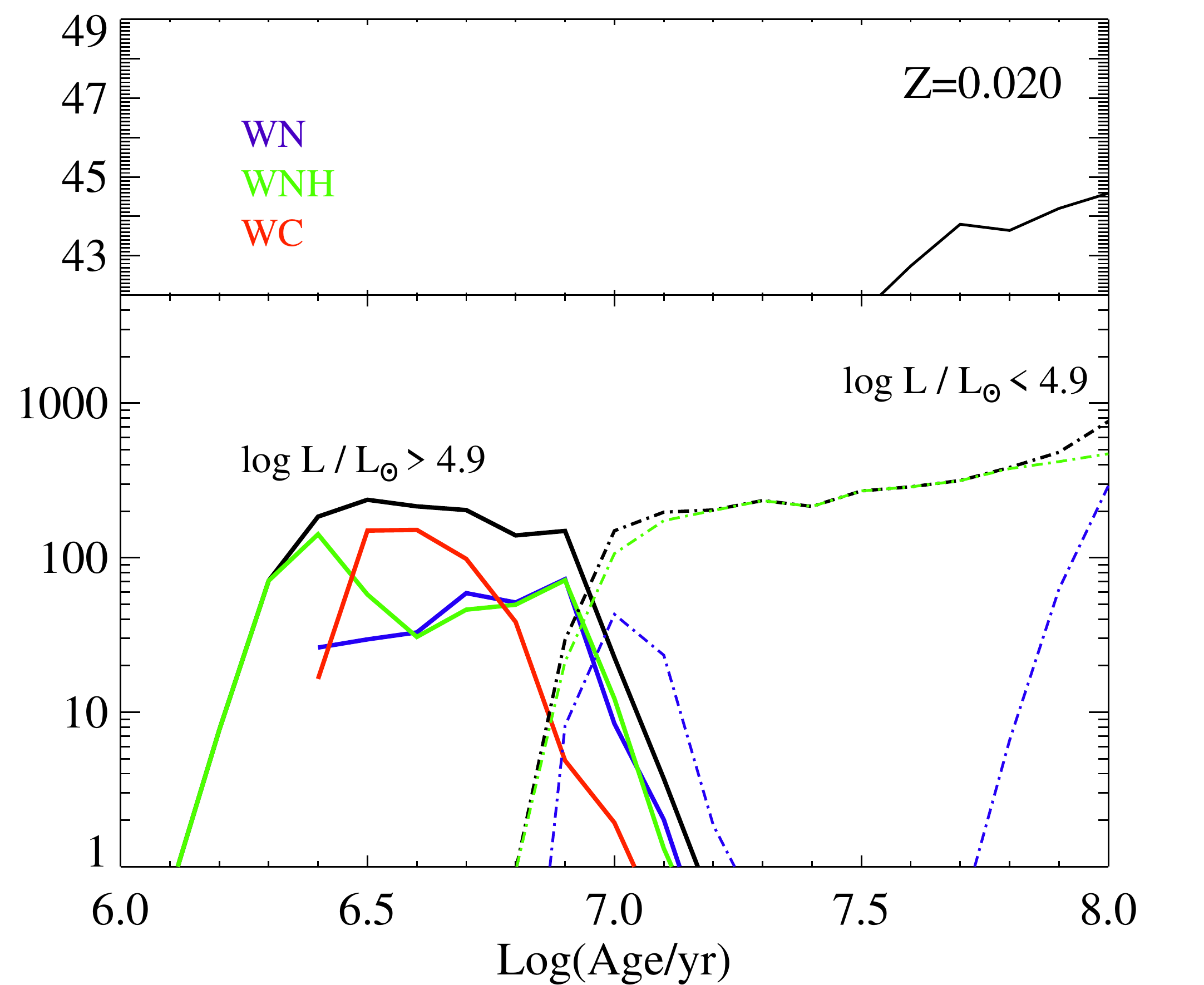}
  \caption{The population of stars with a surface hydrogen mass fraction of $X_\mathrm{surf}<0.4$, i.e. helium stars, as a function of time assuming an initial population of $10^6$\,M$_\odot$, compared to the photon flux Q(O\,VI). Those with very low hydrogen ($X_\mathrm{surf}<0.001$, classic Wolf-Rayet stars) are subdivided by the presence (WC) or absence (WN) of carbon. The population is further subdivided into high luminosity (solid lines) and low luminosity (dot-dash) populations. We show results at the highest and lowest metallicities considered here.}\label{fig:wrevol}
 \end{figure*}

The origin of the short-lived peaks in hard photon production is worth consideration. These are traditionally associated with the evolution of Wolf-Rayet stars. The stripped envelopes and evolved cores of these stars allow them to expand and become luminous while remaining hot and blue. As figure \ref{fig:wrevol} (left panel) demonstrates, the timescales for formation and evolution of these stars are in good agreement with the Q(O\,VI) flux in our models at low metallicity, suggesting that these might be a dominant source of hard ultraviolet emission. It is notable that the dominant contribution at early times is, in fact, from the WNH population of partially-stripped stars which still show measurable surface hydrogen fractions, rather than the most stripped (WN or WC) Wolf-Rayet stars. 

However, an examination of the right-hand panel of figure \ref{fig:wrevol} suggests that care should be taken in such interpretation. A high metallicity starburst will generate as many, or even more, Wolf-Rayet stars as one that is near-primordial, while showing very little hard ionizing flux. In fact, at $Z=0.020$ (Solar metallicity), the entirety of the high energy flux arises from the more numerous but far less luminous helium dwarf population generated by binary interactions of intermediate mass stars. 

However, while Solar-metallicity Wolf-Rayet stars are more numerous, they are also  substantially less luminous than their counterparts at low metallicities. The weak stellar winds associated with low photospheric metal fractions lead to low metallicity stars retaining much of their mass through their evolution which leads to higher core temperatures to achieve hydrostatic equilibrium and thus higher stellar luminosities. For stars that accrete material through binary mass transfer, the weaker wind mass loss also allows these stars to retain more angular momentum and so they experience strong rotational mixing that causes chemically homogenous evolution to occur. These stars have high surface temperatures for their entire hydrogen-burning lifetimes and contribute significantly to the ionizing flux. The factors all result is much higher luminosities.  

\begin{figure*}
    \centering
    \includegraphics[width=0.4\textwidth]{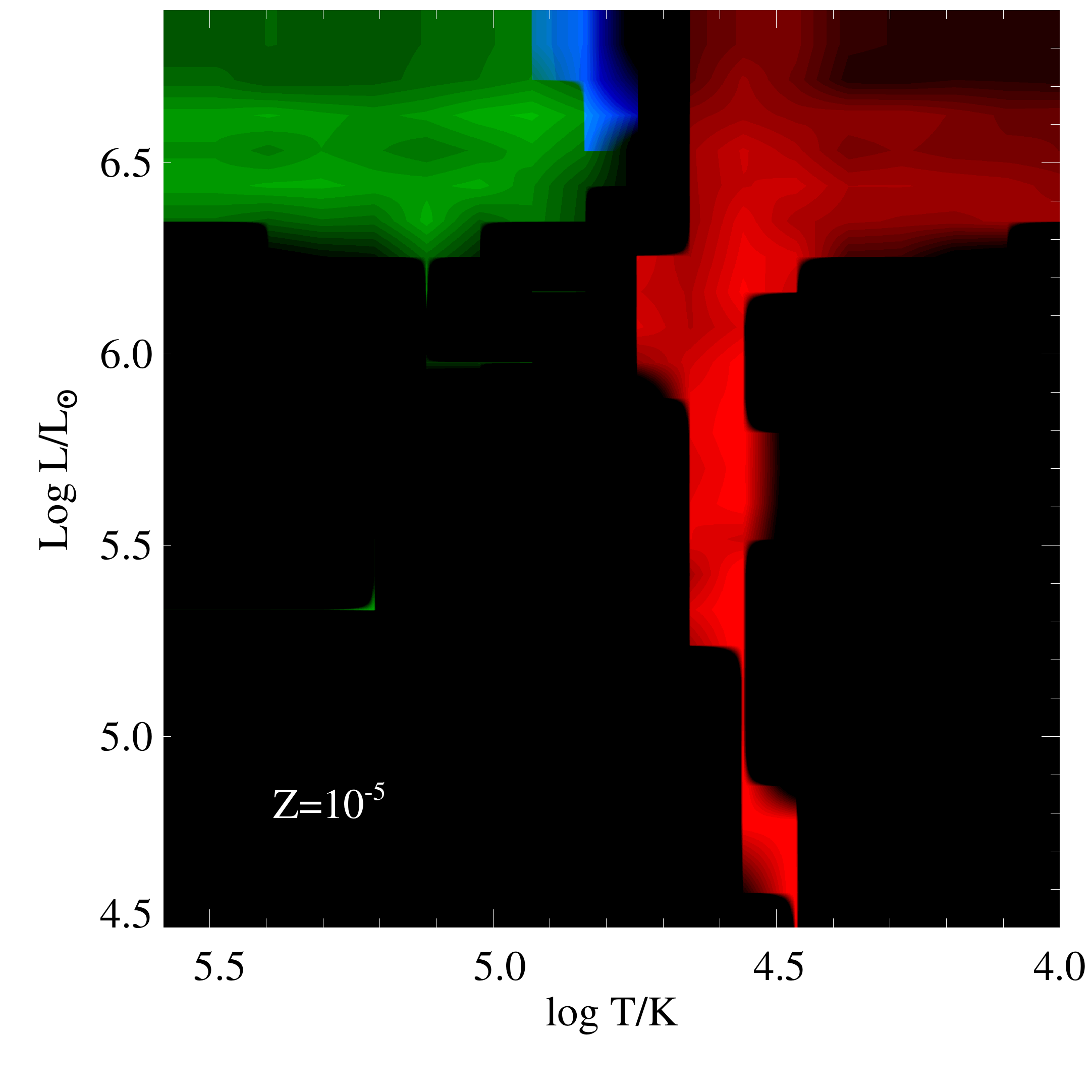}
    \includegraphics[width=0.4\textwidth]{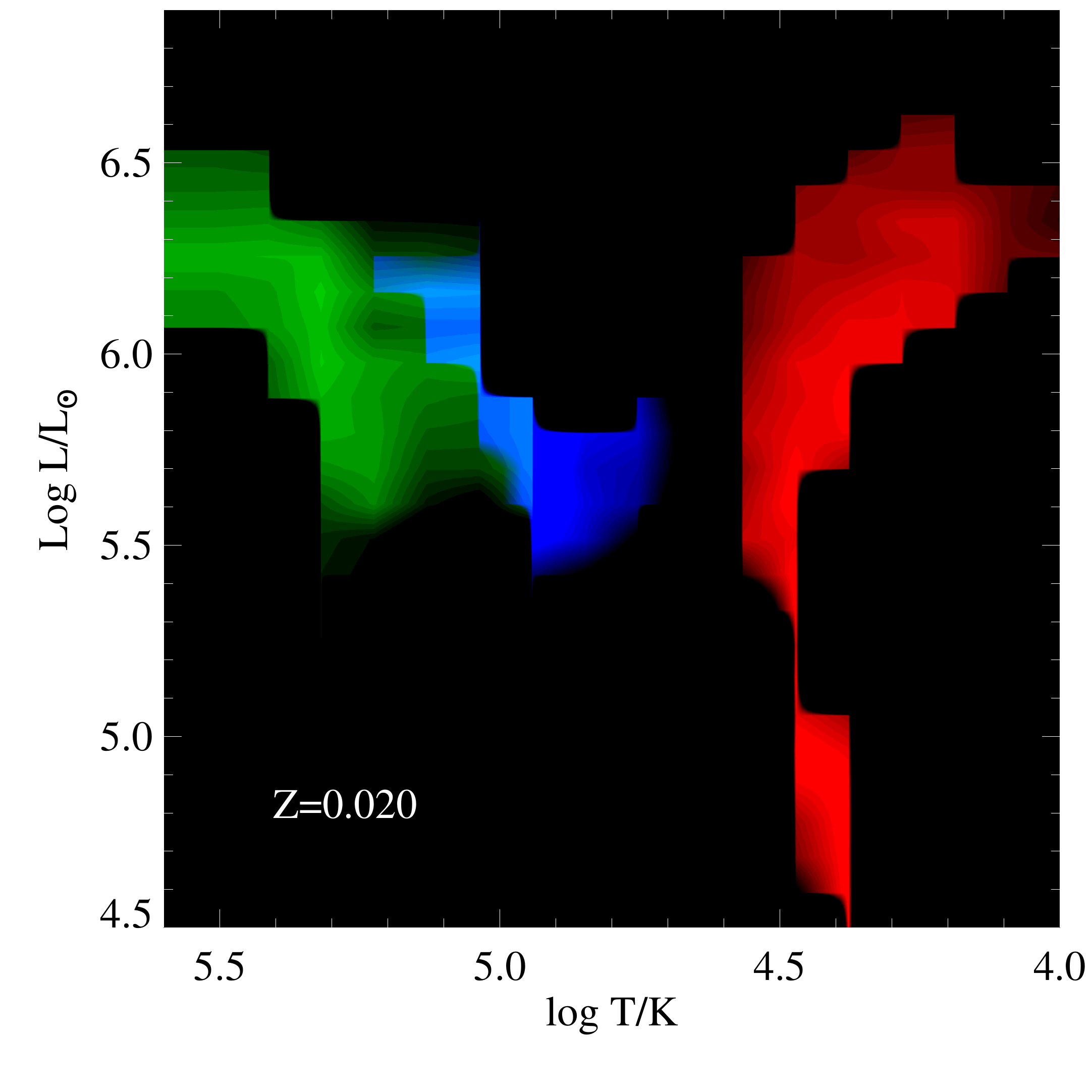}
  \caption{The Herzsprung-Russell diagram for stellar populations at a log(age/years)=6.4, shown at high and low metallicity. Red regions indicate
  stars with a surface hydrogen mass fraction of $X_\mathrm{surf}>0.4$, blue those with $X_\mathrm{surf}<0.001$, and green helium stars with an intermediate surface hydrogen fraction. The same density scaling has been applied to both metallicities. Although the number of hot stars is comparable at $Z=0.020$ to $Z=10^{-5}$, each star is substantially less luminous than its low metallicity counterpart.}\label{fig:wrhrs}
 \end{figure*}

Figure \ref{fig:wrhrs} illustrates this with a traditional Hertzsprung-Russell diagram, indicating the luminosity and temperature distribution of stars at the hard emission peak age of log(age/years)=6.4, as a function of their surface hydrogen mass fraction. The fraction of partially-stripped stars is higher, and the individual stars typically $\sim0.5$\,dex more luminous, at $Z=10^{-5}$ than at $Z=0.020$. At the same time, the metal abundance in the photospheres is lower. As a result, the spectrum of such  starburst populations may be expected to produce substantially weaker traditional "Wolf-Rayet" spectral features (dominated by wind-broadened helium and carbon lines), while still being significant sources of hard ionizing photons. It is unclear whether the Wolf-Rayet atmosphere models currently in use for spectral synthesis fully capture this behaviour at low metallicities.

\begin{figure*}
    \centering
    \includegraphics[width=0.48\textwidth]{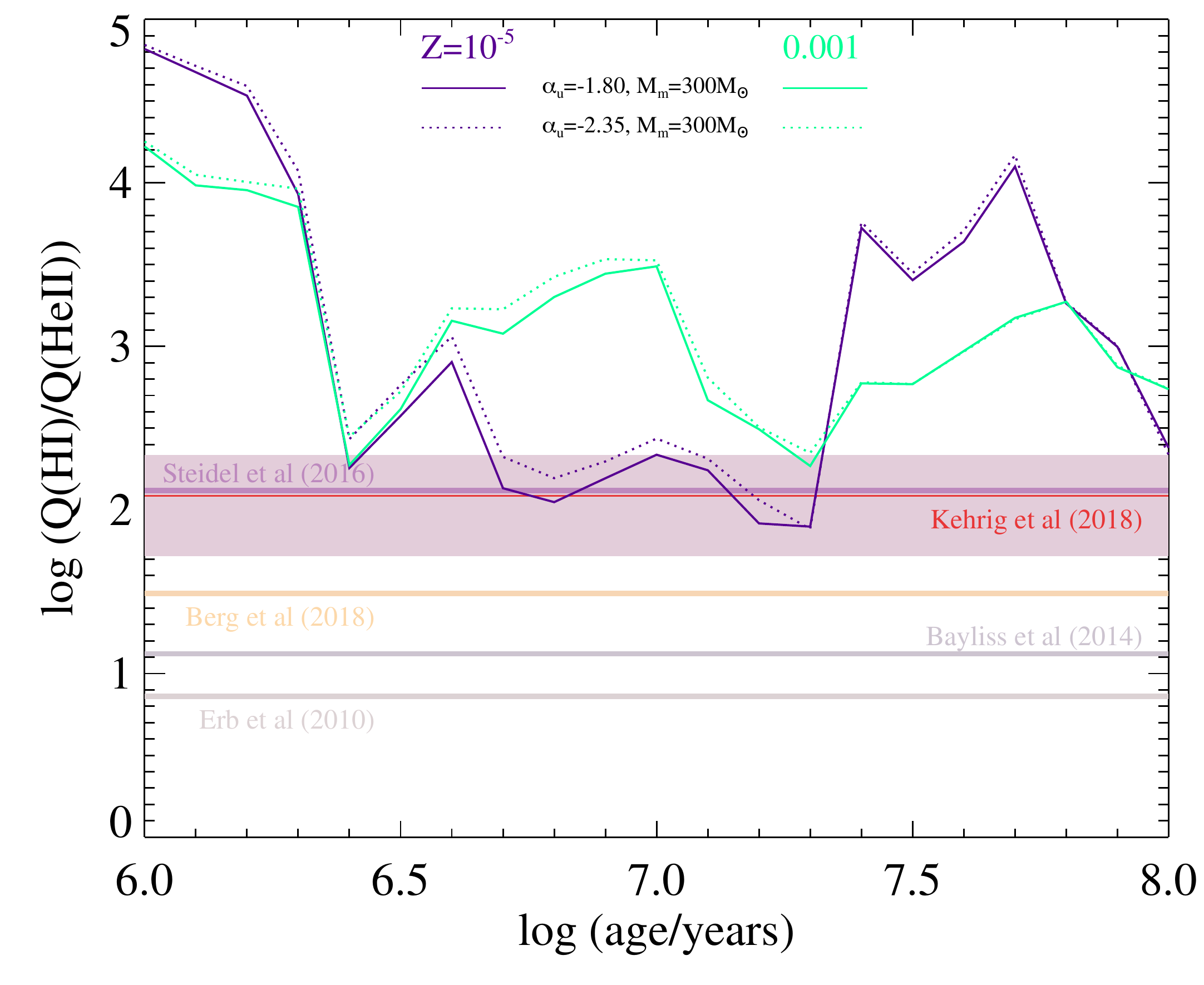}
    \includegraphics[width=0.48\textwidth]{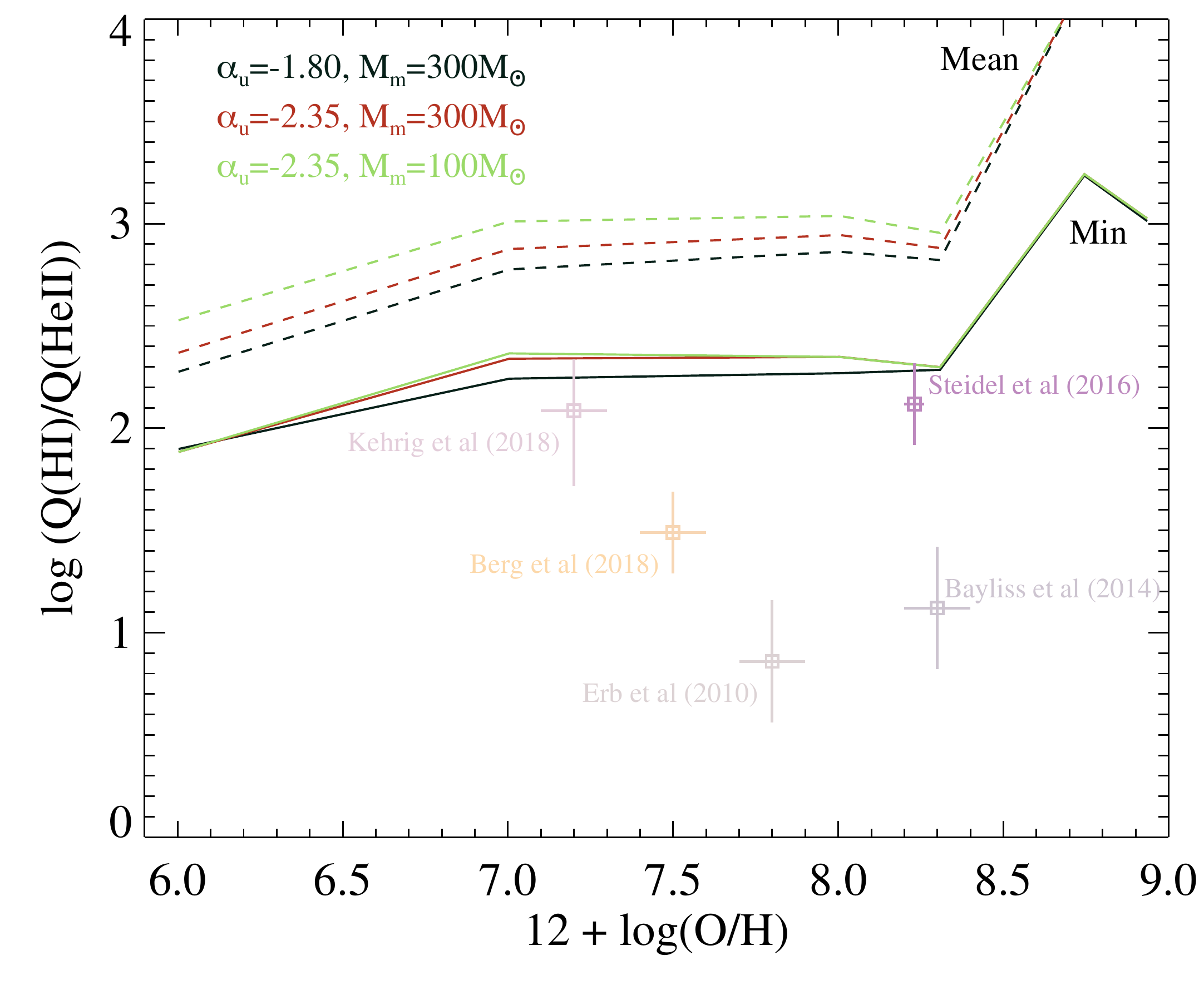}
    \caption{The Q(H\,I)/Q(He\,II) ratios attained by our most extreme models (typically those with $\alpha_m=\alpha_u=-1.8$ and $M_u=300$\,M$_\odot$) compared to those of the BPASS default model ( $\alpha_m=\alpha_u=-2.35$ and $M_u=300$\,M$_\odot$) and observational constraints from the literature. In the left-hand panel we show the time evolution of the photon flux ratio at two metallicities, illustrating the brief lifetime of the hardest ratios. In the right hand panel we show both the hardest ratios attained at a given metallicity and the mean ratio over the first 20\,Myr of stellar evolution. In both panels, we compare to observational data as indicated in the labels and in section \ref{sec:disc}. The filled region in the left hand panel (and error bars on the relevant point in the right hand panel) indicates the range of values measured for different regions in the resolved galaxy SBS\,0335-052E \citep[$D=54$\,Mpc,][]{2018MNRAS.480.1081K}.}
    \label{fig:obsdat}
\end{figure*}

As we have demonstrated, IMF variations can substantially affect the ionizing flux. It is interesting to note, however, that the variations in the ionizing spectrum are somewhat mitigated by the fact that the entire spectrum is responding in a similar way to initial mass function.  As figure \ref{fig:obsdat} illustrates, even the most extreme ionizing populations in the grid we have calculated reach peak Q(He\,II)/Q(H\,I) ratios (i.e. spectral hardness measures) which only vary by a few tenths of a dex from those of the default BPASS initial mass function at the same metallicity.

This presents a challenge. The increasing sample of ionizing galaxies in the literature showing evidence for a hard ionizing photon spectrum motivated this study and constraints on the spectral properties of these sources are also shown in figure \ref{fig:obsdat}. We have restricted our analysis to a representatitive sample of extreme sources with measurements of both the H$_\beta$ and He\,II 1640\AA\ or 4686\AA\ recombination lines. While other lines, including [C\,IV] 1540, C\,III] 1909 and O\,III] 1665 are also indicative of hard ionizing fields, they are also highly sensitive to nebular gas conditions complicating interpretation of line strength. By contrast, the recombination lines of hydrogen and helium are only weakly sensitive to electron density and temperature, such that,
\begin{equation*}
\hspace{0.5cm} \frac{Q(H\,I)}{Q(He\,II)} = \frac{F_H\,\lambda_H}{F_{He}\,\lambda_{He}}\frac{\alpha^H_B}{\alpha^H_\mathrm{eff}} \frac{\alpha^{He}_\mathrm{eff}}{\alpha^{He}_B}\,,
\end{equation*}
where $F_X$ and $\lambda_X$ are the (extinction and lensing corrected) flux and wavelength of the lines used, $\alpha^X_B$ are the case B recombination coefficients and $\alpha^X_\mathrm{eff}$ are the effective recombination coefficients relevant to each specific line. For the observational data in figure \ref{fig:obsdat} we calculate $\alpha^X_B$ and $\alpha^X_\mathrm{eff}$ for $N_e=10^3$\,cm$^{-3}$ and $T_e=15,000$\,K, using a simple linear interpolation between the values given by \citet{2006agna.book.....O} for each line. The electron density and temperature of extreme ionizing sources appear to vary from source to source, from region to region and even between different tracers from species to species within a source. The electron density and temperature assumed here is motivated by observations of SL2S\,J021737-051329 at $z=1.8$ \citep{2018ApJ...859..164B} and by the mean properties of stacked $z\sim2.3$ Lyman break galaxies \citep{2016ApJ...826..159S}. They may be slightly higher than are appropriate for Q2343-BX418 \citep[$z=2.3$,][]{2010ApJ...719.1168E} and  SGAS\,J105039.6+001730 \citep[$z=3.6$,][]{2014ApJ...790..144B}, but the inferred values for Q(H\,I)/Q(He\,II) are only weakly dependent on this. For SBS\,0335-052E \citep[$D=54$\,Mpc,][]{2018MNRAS.480.1081K} we use the values provided by the authors in their table\,1 for both the range of ratios in spatially resolved regions (shown as a shaded region) and the integrated value for the whole galaxy (shown by a line). Where the observational uncertainties on line flux are small, we apply a minimum uncertainty in the photon flux ratio of 0.2\,dex, accounting for the possible effects of extinction, lensing corrections and uncertainty on the recombination coefficients. None of these sources show any significant evidence for AGN activity.

As the figure makes clear, even our most extreme physically-motivated IMF stellar population synthesis models are unable to reproduce the high line ratios (and hence inferred photon flux ratios) observed in some of the most extreme known systems at either high or low redshift.  Although the highest flux ratios reached by our models may be consistent with the observational data for SBS\,0335-052E and the $z\sim2.3$ Lyman break galaxy stack, these will only be attained for brief periods, or at the very lowest metallicities we consider. If the mean ratio observed over a 20\,Myr period is considered instead, the requirement for exceptionally low metallicities (inconsistent with the observed nebular gas) becomes still stronger.  For other objects at $z\sim2-4$, BPASS stellar spectrum models are simply unable to reproduce the line ratios, regardless of assumed IMF. Indeed, even a population comprised entirely of 300\,M$_\odot$ stars and their binary companions never drops below Q(H\,I)/Q(He\,II)=1.3, given our stellar evolution and atmosphere prescriptions. Hence a stochastic IMF sampling heavy in very massive stars cannot reproduce the observational data. This strongly implies that additional sources of hard ionizing photons need to be considered when analysing highly ionized galaxies, either in the local Universe or at high redshift. It is likely that the source of these hard ionizing photons arises from accretion around compact objects. The open question is whether it arises from acretion onto neutron stars and stellar mass black holes in X-ray binaries \citep{2011A&A...528A.149M,2012MNRAS.423.1641J} or supermassive black holes at the center of galaxies acting as AGN \cite{2015ARA&A..53..365N}, neither of which are currently included within the BPASS models.

\begin{figure*}
    \centering
    \includegraphics[width=0.48\textwidth]{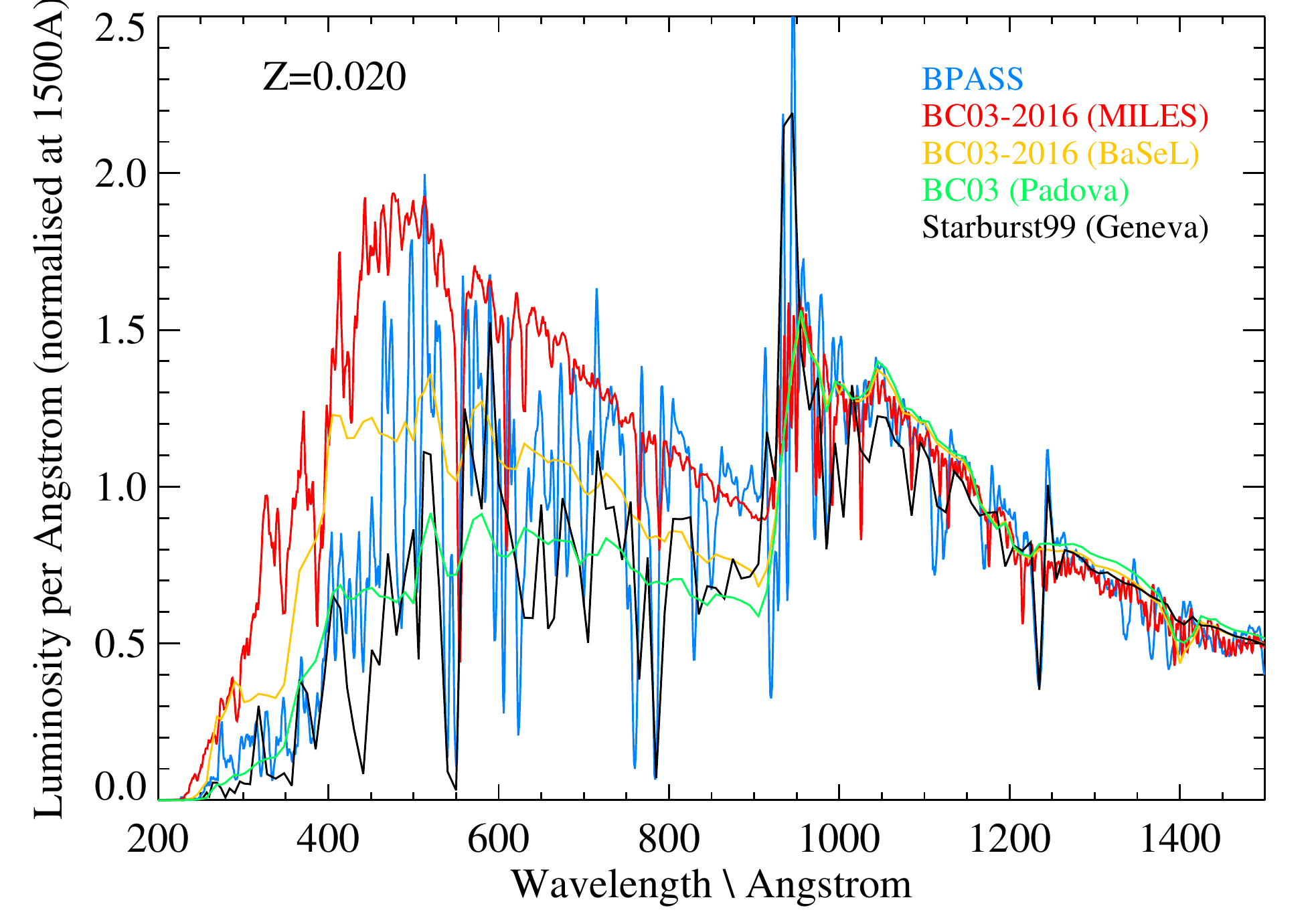}
    \includegraphics[width=0.48\textwidth]{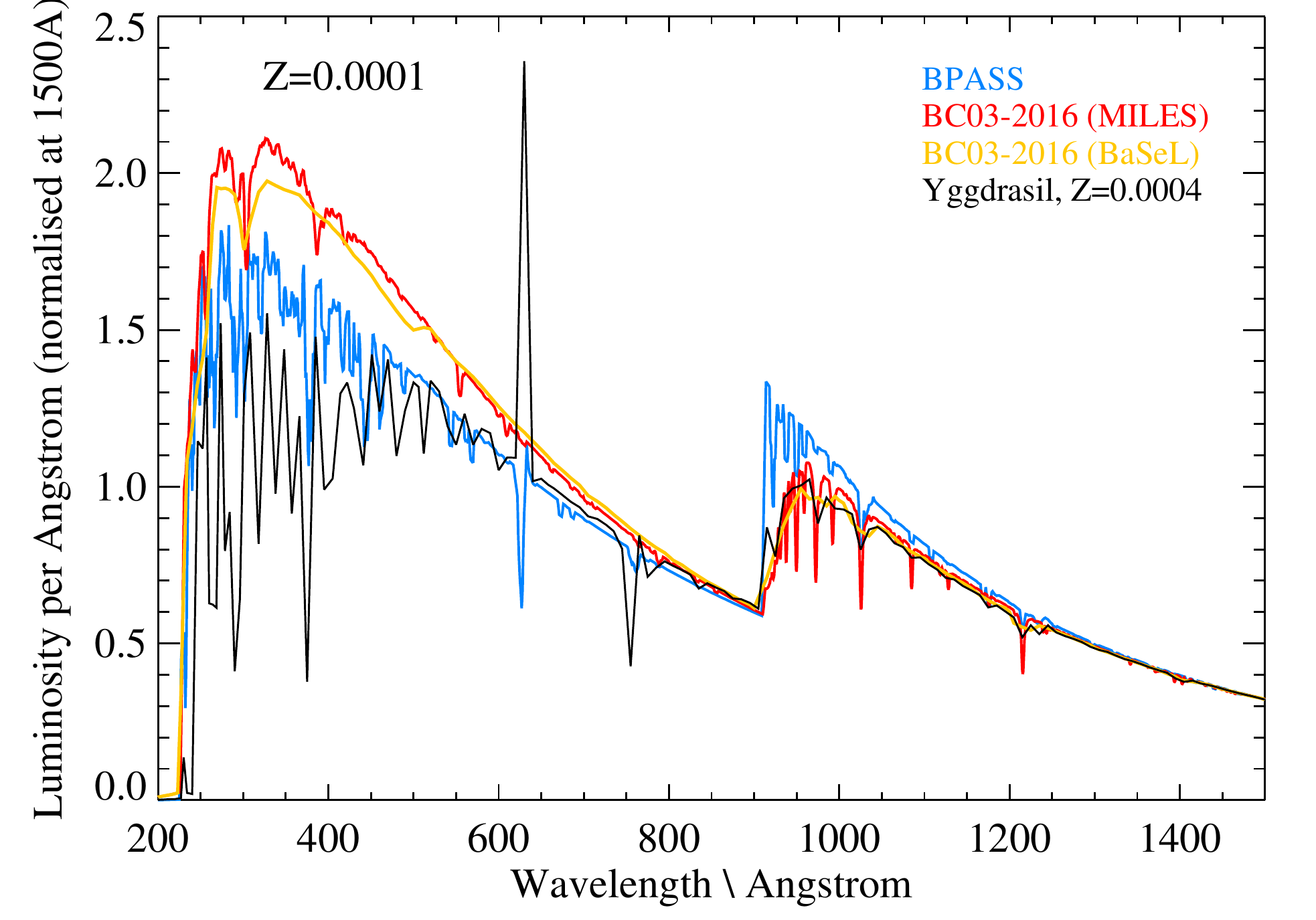}
    \caption{The uncertainty in the extreme ultraviolet spectrum of a 1\,Myr starburst as predicted by a range of stellar population synthesis codes, given the same two metallicities and ($\alpha_u=\alpha_m=-2.35$) initial mass function, but different stellar evolution and atmosphere input models. High resolution models have been slightly smoothed for display purposes. Model references: BPASS \citep{2017PASA...34...58E}, BC03 \citep{2003MNRAS.344.1000B}, Starburst99 \citep{1999ApJS..123....3L,2011ascl.soft04003L}, Yggdrasil \citep{2011ApJ...740...13Z}. The 2016 version of the BC03 models has not been fully described in the refereed literature but some details are to be found in \citet{2016MNRAS.462.1415C} and \citet{2016MNRAS.457.4296W}.}
    \label{fig:seds}
\end{figure*}

This analysis has necessarily focussed on the output of just one stellar evolution and population synthesis code, the assumptions and methodology of which have been clearly described in the literature. However given the range of such models in the literature, it is appropriate to consider whether an alternate choice of stellar population synthesis model might yield a different conclusion. All model sets face two key questions: which stellar evolution tracks to employ, and what stellar atmosphere models to match them to. The former question includes uncertainties such as the impact of binary evolution or rotation, while the latter incorporates our uncertainty regarding radiative transfer through stellar envelopes as well as the difficulty of testing or calibrating predictions for the extreme ultraviolet (which is effectively unobservable due to line-of-sight absorption by interstellar hydrogen). The degree of uncertainty in the extreme ultraviolet emission of a population even at a main sequence age of 1\,Myr is demonstrated in figure \ref{fig:seds}. As the figure demonstrates BPASS models produce substantially more ionizing flux than the older Starburst99 \citep[Geneva, non-rotating,][]{1999ApJS..123....3L,2011ascl.soft04003L} and BC03 \citep[galaxev,][]{2003MNRAS.344.1000B} population synthesis model at Solar metallicity. The 2016 release of the BC03 models presents a mixed picture. Both when using a BaSeL stellar atmosphere grid, and when using the semi-empirical MILES stellar spectra grid, the BC03-2016 model spectra extend to shorter wavelengths than BPASS at Solar metallicity, but the flux normalisation in the ionizing regime appears to be strongly dependent on the stellar atmosphere library in use, with the MILES library producing a calibration similar to BPASS just below the Lyman limit but failing to match the flux predicted by both BPASS and Starburst99 between 912 and 1216\,\AA, and exceeding both shortwards of 500\AA. The origin of the very blue spectrum at Solar metallicity in the BC03-2016 models is unclear (the details of implementation in this code have not appeared in the literature) but is likely associated with a boost in the number or luminosity of Wolf-Rayet spectra incorporated by the population synthesis. The same trends are seen at lower metallicities. At $Z=10^{-4}$ the extreme ultraviolet spectrum is unconstrained by observations, and the implementation favoured by BC03-2016 is significantly bluer than that of BPASS for the same input IMF. The only other SPS model which is available at this very low metallicity is the Yggdrasil model of \citep{2011ApJ...740...13Z} which has $Z=4\times10^{-4}$. This is very similar to the $Z=10^{-4}$ model of BC03-2016 longwards of the Lyman limit, but more similar to BPASS below it and deficient in photons with $\lambda<500$\,\AA.

\begin{figure}
    \centering
    \includegraphics[width=0.48\textwidth]{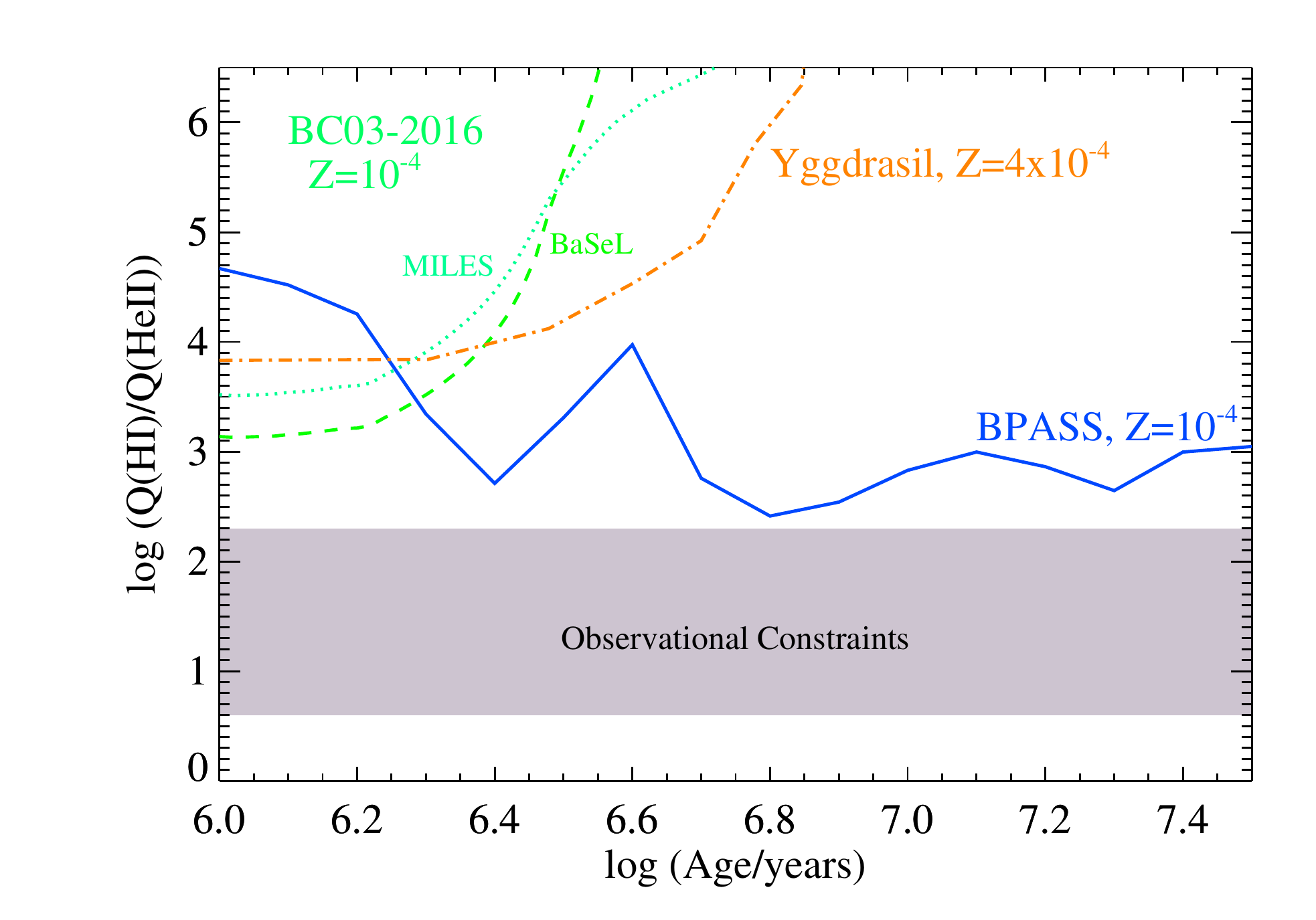}
    \caption{The effect of binary stellar pathways on the evolution of the Q(H\,I)/Q(He\,II) ratios attained by stellar population synthesis models at low metallicities. The tracks show the ionizing photon flux ratio with time for the BPASS and BC03-2016 models at $Z=1\times10^{-4}$ and the Yggdrasil model at $Z=4\times10^{-4}$. The region spanned by the observational constraints in figure \ref{fig:seds} is shaded. While the single-star evolution models begin life bluer at these metallicities, binary pathways generate hard ionizing flux at later ages.}
    \label{fig:spscomp}
\end{figure}

However, as figure \ref{fig:spscomp} demonstrates, the behaviour of the SPS models at effectively zero age may not be the key diagnostic in interpreting the spectra of high redshift stellar populations. Both the low metallicity BC03-2016 models and the Yggdrasil models evolve rapidly, with strong Q(He\,II) emission fading away within 2-3\,Myr after the stars are formed. By contrast, binary evolution pathways generate more hot stars at later ages, and as a result the BPASS models maintain strong Q(He\,II) emission relative to Q(H\,I) over timescales of 30\,Myr or longer. While all the models fail to match the values inferred from the observational constraints, timescale arguments suggest that we are unlikely to be catching so many systems at ages $<3$\,Myr, somewhat favouring the binary evolution models.

\section{Conclusions}\label{sec:conc}

Our main conclusions can be summarised as follows:

\begin{enumerate}
    \item Young starbursts at low metallicity can generate the high energy photons (shortwards of 228\AA\ and 89.8\,\AA) required to power emission or absorption features in the He\,II and O\,VI lines.
    \item The photon flux at these energies is 2.5 orders of magnitude lower than that shortwards of 912\AA\ for Q(He\,II), and five orders of magnitude lower for Q(O\,VI), even when at its peak and at very low metallicities.
    \item Hard ionizing radiation, particularly Q(O\,VI), is associated with the most massive stars in the population ($M>100$\,M$_\odot$), and so is short-lived and boosted by shallower power law initial mass functions or higher mass IMF cut-offs.
    \item The probability of these very massive stars occurring in a starburst with a low total mass is $<<1$, meaning the fluxes in Q(He\,II) and Q(O\,VI) are highly susceptible to stochastic variation, even in galaxy-wide starbursts.
    \item Even the most extreme initial mass functions in a grid motivated by observations of stellar masses in the local Universe are unable to reproduce the low Q(H\,I)/Q(He\,II) ratios observed in extreme ionizing sources in either the local or distant Universe.
\end{enumerate}

The analysis presented here demonstrates the range of hard photon emission rates that result from plausible assumptions regarding the stellar initial mass function. We caution that detailed interpretation of features in individual galaxy spectra is inevitably going to be subject to uncertainties in the IMF of its contributing starbursts. We remind the community that the initial mass function is fundamentally a statistical construct, and that stellar population synthesis models are most effective when considering entire galaxy populations rather than individual objects. Nonetheless, it seems likely that for some of the hardest ionization potentials inferred for galaxies in the distant Universe, IMF variations are inadequate to explain the observed emission, and additional sources of short-wavelength photons beyond those from stellar photospheres are required.

\begin{acknowledgements}
BPASS would not be possible without the computational resources of the University of Auckland's NeSI Pan Cluster and the University of Warwick's Scientific Computing Research Technology Platform (SCRTP). New Zealand's national facilities are provided by the NZ eScience Infrastructure and funded jointly by NeSI's collaborator institutions and through the Ministry of Business, Innovation \& Employment's Research Infrastructure programme.  We thank both the University of Warwick and the University of Auckland for travel support. ERS thanks John Chisholm and Steve Wilkins for useful discussions. We also thank Polychronis Papaderos and the other organisers of the ``Escape of Lyman-$\alpha$ from Galactic Labyrinths'' conference (OAC, Crete, 2018) for the stimulus that motivated this work.
 \end{acknowledgements}

%-------------------------------------------------------------------

   \bibliographystyle{aa} % style aa.bst
   \bibliography{imf_nphot} % your references Yourfile.bib

\end{document}